\documentclass[letter,utf8]{aa}

\usepackage{graphicx}
\usepackage{xspace}
\usepackage{color}
\usepackage{natbib}
\usepackage{ulem}
\usepackage{longtable}
\usepackage{multirow}
\usepackage{lscape}
\usepackage{tikz}
\usepackage{supertabular}
\usepackage{circledsteps}
\usepackage{txfonts}
\usepackage[colorlinks=true,citecolor=blue,linkcolor=blue]{hyperref}
\bibpunct{(}{)}{;}{a}{}{,} % to follow the A&A style

\newcommand{\mission}[1]{\textit{#1}}
\newcommand{\erasst}{eRASSt~J045650.3$-$203750\xspace}
\newcommand{\jsrc}{J0456$-$20\xspace}
\newcommand{\asko}{ASASSN-14ko\xspace}
\newcommand{\hlx}{HLX-1\xspace}
\newcommand{\gsn}{GSN~069\xspace}

\newcommand{\rxj}{RX~J133157.6$-$324319.7\xspace}

\newcommand{\ptde}{\textit{p}TDE\xspace}
\newcommand{\ptdes}{\textit{p}TDEs\xspace}

\newcommand{\trecur}{T_\text{recur}}

\newcommand{\fxfaint}{${P}_\text{X, faint}$\xspace}
\newcommand{\fxdrop}{${P}_\text{X, drop}$\xspace}
\newcommand{\fxrise}{${P}_\text{X, rise}$\xspace}
\newcommand{\fxplat}{${P}_\text{X, plat}$\xspace}

\newcommand{\unitflux}{\text{erg\,cm}^{-2}\,\text{s}^{-1}}

\newcommand{\unitlumi}{\text{erg\,s}^{-1}}
\newcommand{\msun}{\mathrm{M}_\odot}

\newcommand{\mbh}{M_\text{BH}}

\newcommand{\sflux}{$f_\text{X, soft}$\xspace}

\newcommand{\mpl}{$\text{M}_\text{pl}$\xspace}

\newcommand{\mmcd}{$\text{M}_\text{mcd}$\xspace}

\begin{document}

   \title{Rapid evolution of the recurrence time in the repeating partial tidal disruption event \erasst}

   \author{Zhu Liu\inst{\ref{ins:mpe}}
           \and Taeho Ryu\inst{\ref{ins:mpa}}
           \and A. J. Goodwin\inst{\ref{ins:curtinu}}
           \and A. Rau\inst{\ref{ins:mpe}}
           \and D. Homan\inst{\ref{ins:aip}}
           \and M. Krumpe\inst{\ref{ins:aip}}
           \and A. Merloni\inst{\ref{ins:mpe}}
           \and I. Grotova\inst{\ref{ins:mpe}}
           \and G. E. Anderson\inst{\ref{ins:curtinu}}
           \and A. Malyali\inst{\ref{ins:mpe}}
           \and J. C. A. Miller-Jones\inst{\ref{ins:curtinu}}
          }

   \institute{
              %{\bf Co-author list is currently in alphabetical order, and will be determined before submission. Please let me know or add them in, if you find someone are not in the list.} \label{expl}
              Max-Planck-Institut für extraterrestrische Physik,
                   Gie{\ss}enbachstra{\ss}e 1, 85748 Garching, Germany\label{ins:mpe}\\
              \email{liuzhu@mpe.mpg.de}
              \and Max-Planck-Institut für Astrophysik,
                    Karl-Schwarzschild-Str. 1, 85748 Garching, Germany\label{ins:mpa}
              \and International Centre for Radio Astronomy Research,
                   Curtin University, GPO Box U1987, Perth, WA 6845, Australia\label{ins:curtinu}
              \and Leibniz-Institut für Astrophysik Potsdam,
                   An der Sternwarte 16, 14482 Potsdam, Germany\label{ins:aip}
             }

   \date{Received xxx xx, xxxx; accepted xxx xx, xxxx}

   \abstract{In this letter, we present the results from further X-ray and UV observations of the nuclear transient \erasst (hereafter \jsrc). We detected five repeating X-ray and UV flares from \jsrc, making it one of the most promising repeating partial tidal disruption event (\ptde) candidates. More importantly, we also found rapid changes in the recurrence time $\trecur$ of the X-ray flares by modelling the long-term X-ray light curve of \jsrc. $\trecur$ first decreased rapidly from about 300~days to around 230~days. It continued to decrease to around 190~days with an indication of a constant $\trecur$ evidenced from the latest three cycles. Our hydrodynamic simulations suggest that, in the repeating \ptde scenario, such rapid evolution of $\trecur$ could be reproduced if the original star is a $1\,\msun$ main-sequence star near the terminal age and loses nearly 80--90\% of its mass during the initial encounter with a supermassive black hole (SMBH) of mass around $10^5~\msun$. The inferred mass loss of 0.8--0.9~$\msun$ is higher than the estimated value of around 0.13~$\msun$ from observation, which could be explained if the radiation efficiency is low (i.e. $\ll0.1$). Our results indicate that repeating \ptdes could be effective tools to explore the dynamics around supermassive black holes beyond our own Galaxy.}
% 5 {} token are mandatory

   \titlerunning{Rapid recurrence time evolution in \jsrc}
   \authorrunning{Liu et al.}
   \maketitle
%
%-------------------------------------------------------------------

\section{Introduction}

Tidal disruption events (TDEs) are typically considered one-off events where a star is completely destroyed by a supermassive black hole (SMBH) at the first pericenter passage. However, theoretical calculations and numeric simulations have shown that a partial TDE (\ptde) can also occur \citep[e.g.][]{guillochon_etal2013, ryu_etal2020c}. In a \ptde, the star loses only a fraction of its mass and survives its first encounter with the SMBH. If the star is initially in a bound orbit with low eccentricity, it is expected to generate repeating flares \citep[e.g.][]{hayasaki_etal2013, ryu_etal2020c, nixon_etal2022, cufari_etal2022, cufari_etal2023, melchor_etal2024}. Because the encounter cross sections of \ptdes are generally larger than or comparable to those of full TDEs, the rate of \ptdes is expected to be larger or comparable to full TDEs \citep{Krolikc+2020,Bortolas+2023}. Repeating \ptdes are particularly interesting as they may be effective probes to explore stellar dynamics around SMBHs beyond our own Galaxy and as they provide ideal laboratories to study the accretion processes in SMBHs.

Only a few repeating \ptde candidates have been reported so far (e.g. \asko, \citealt{payne_etal2021}; \hlx\footnote{Note that \hlx is believed to be an intermediate mass black hole (IMBH) with mass of $\sim10^{4-5}\msun$ \citep{davis_etal2011, webb_etal2012}}, \citealt{webb_etal2023}; \erasst, hereafter \jsrc, \citealt{liu_etal2023}; \rxj, \citealt{malyali_etal2023}; AT2018fyk, \citealt{wevers_etal2023}). Among these, \jsrc is one of the best-studied repeating nuclear transients discovered in a quiescent galaxy ($z=0.077$). \citet{liu_etal2023} reported the detection of three repeated X-ray and UV flares from \jsrc. In particular, the profiles of the X-ray flares are similar and can be characterised by four distinctive phases: an {X-ray rising} phase (\fxrise) leading into an {X-ray plateau} phase (\fxplat), which is terminated by a rapid {X-ray drop} phase (\fxdrop), and followed by an {X-ray faint} state (\fxfaint). These results provide strong evidence that \jsrc is a repeating nuclear transient, making \jsrc one of the most promising \ptde candidates.

Quasi-period eruptions (QPEs) are a class of recurring X-ray flares found in galactic nuclei, with periods less than one day \citep{miniutti_etal2019, giustini_etal2020, arcodia_etal2021}. Their origin remains elusive, though recent studies indicate a potential link to \ptdes (e.g. \gsn, \citealt{miniutti_etal2023}; \rxj, \citealt{malyali_etal2023}). The discovery of Swift~J0230, which exhibits QPE-like behaviours with a period of around 22~days \citep{evans_etal2023, guolo_etal2023}, further strengthens the link between repeating \ptdes and QPEs. It is thus interesting to study the evolution of $\trecur$ of the flares for repeating \ptdes. However, substantial evolution in $\trecur$ has only been reported in \asko and \hlx. A period derivative of $-0.0026\pm0.0006$, with a period of $115.2^{+1.3}_{-1.2}~$days, has been reported for \asko \citep{payne_etal2022}. \hlx initially shows quasi-periodic X-ray outbursts spaced by about $1$~yr between 2009 and 2012. The $\trecur$ then started to increase until 2018, after which no X-ray outbursts were detected \citep{godet_etal2014,webb_etal2023}. Long-term multi-wavelength observations on a larger sample of repeating \ptdes are required to fully understand the evolution of $\trecur$ in these objects and potentially provide further evidence to confirm the connection between repeating \ptdes and QPEs.

\begin{figure*}
    \centering
    \includegraphics[width=\textwidth]{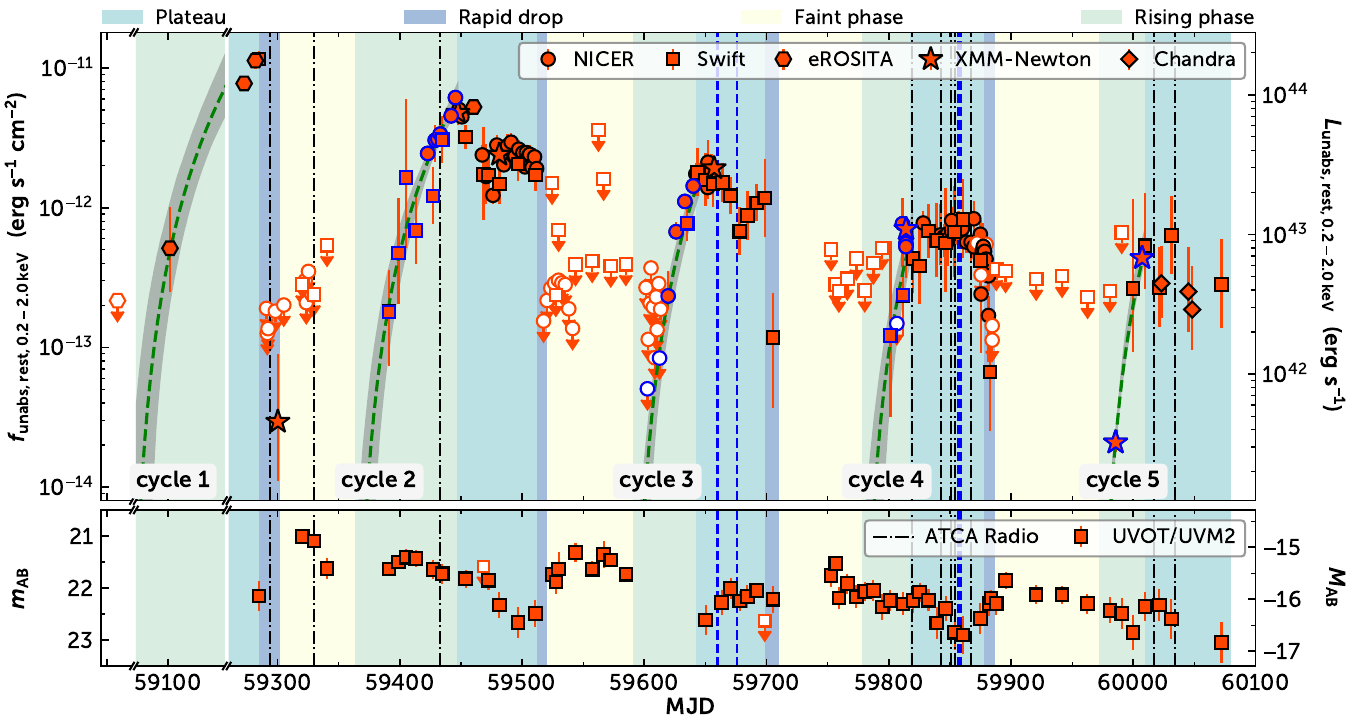}
    \caption{Long-term X-ray and UV light curves for \jsrc. The coloured regions represent the four phases: the plateau phase (\fxplat, light cyan), the rapid drop phase (\fxdrop, light blue), the faint phase (\fxfaint, light yellow), and the rising phase (\fxrise, light green). \textit{Upper panel:} Red points with error bars are the unabsorbed rest-frame $0.2-2.0\,\text{keV}$ X-ray light curve from eROSITA (hexagons), \mission{Swift}/XRT (squares), \mission{NICER} (circles), \mission{XMM-Newton} (stars), and \mission{Chandra} (diamonds). The error bars indicate 90\% uncertainties. The points with downward arrows represent the $3\sigma$ flux/luminosity upper limits. The points with blue colour are the data used to model the profiles of the \fxrise phase. The green dashed lines show the best-fitting power-law model for the five X-ray rising phases. The grey shaded regions mark the $1\sigma$ uncertainty of the model. \textit{Bottom panel:} UV light curve from \mission{Swift}/UVOT UVM2 (red squares). The error bars mark the $1\sigma$ uncertainties. Squares with downward arrows indicate 3$\sigma$ upper limits. The vertical lines mark the dates of the ATCA radio observations (black dashed-dotted: non-detections; blue dashed: detections).}
    \label{fig:multi_lc}%
\end{figure*}

In this letter, we report the discovery of the rapid evolution of $\trecur$ in \jsrc. A detailed analysis of the latest X-ray and UV data reveals five X-ray and UV flares in \jsrc (marked as cycle 1--5 in Fig.~\ref{fig:multi_lc}). These results confirm that \jsrc is indeed a repeating nuclear transient. In addition, they also suggest an initially rapid decrease in $\trecur$ by more than two months between cycles 1 and 2, slowing to around $40~$days between cycles 2 and 3. $\trecur$ reaches an almost constant value of around $190~$days in the latest three cycles. This paper is structured as follows. In Sect.\,\ref{sec:multi_band}, we present the multi-wavelength data reduction. X-ray spectral and light curve modelling are presented in Sect.\,\ref{sec:results}. Finally, we discuss and summarise our results in Sects.\,\ref{sec:discussion} and \ref{sec:summary}. Throughout this paper, we adopt a flat $\Lambda$CDM cosmology with $H_0=67.7\,\text{km\,s}^{-1}\,\text{Mpc}^{-1}$ and $\Omega_m=0.308$ \citep{planck_etal2020}. Therefore $z=0.077$ corresponds to a luminosity distance of $D_\text{ld}=360\,\text{Mpc}$. All magnitudes will be reported in the AB system (not corrected for Galactic extinction). All the quoted uncertainties correspond to the 90\% confidence level, unless specified otherwise.

%--------------------------------------------------------------------
\section{Observations and Data Reduction}\label{sec:multi_band}
In this paper, we analysed all observations after MJD~59720. The reader should refer to \citet{liu_etal2023} for the details of data analysis before this date.

\subsection{\mission{XMM-Newton}\label{subsec:multi_xmm}}
A pre-approved \mission{XMM-Newton} target of opportunity (ToO) observation was performed on 2022 Aug 23 (hereafter X5). In addition, two \mission{XMM-Newton} Director's Discretionary Time (DDT) observations were performed on 2023 Feb 10 and Mar 4 (hereafter X6 and X7, respectively). \jsrc was detected in all new \mission{XMM-Newton} observations.

The Observation Data Files (ODFs) for each observation were reduced using the \mission{XMM-Newton} Science Analysis System software (\textsc{SAS}, version \texttt{19.1}, \citealt{gabriel_etal2004}) with the latest calibration files. For each observation, the \textsc{SAS} tasks \texttt{emchain} and \texttt{epchain} tasks were used to generate the event lists for the European Photon Imaging Camera (EPIC) MOS \citep{turner_etal2001} and pn \citep{struder_etal2001} detectors, respectively. High background flaring periods were identified and filtered from the event lists. For all the EPIC images, a circular region with a radii of $30, 20, 25''$ was chosen as the source region for X5, X6, and X7, respectively. A source-free annulus region with an inner radius of $50''$ and an outer radius of $100''$ was chosen as the background region for all MOS observations. The background for the pn camera was extracted from a circular region with a radius of $60''$ centred at the same CCD read-out column as the source position for all observations. X-ray events with pattern $\leq12$ for MOS and $\leq4$ for pn were selected to extract the X-ray spectra. We used the tasks \texttt{rmfgen} and \texttt{arfgen} to generate the response matrix and ancillary files, respectively. The X-ray spectra were rebinned to have at least 1 count per bin.

\subsection{\mission{Swift} observations\label{subsec:multi_sw}}

The XRT online data analysis tool\footnote{\url{http://www.swift.ac.uk/user_objects}} \citep{evans_etal2009} was used to check whether the source was detected for each observation. It was also used to generate the X-ray spectra for observations in which \jsrc was detected and to calculate the $3\sigma$ count rate upper limits for non-detections. The X-ray spectra were rebinned to have at least 1 count in each bin.

The \mission{Swift}/UVOT data were reduced using the UVOT analysis pipeline provided in \textsc{HEASoft} (version \texttt{6.31}) with UVOT calibration version \texttt{20201215}. Source counts were extracted from a circular region with a radius of $5''$ centred at the source position. A $20''$ radius circle from a source-free region close to the position of \jsrc was chosen as the background region. The task \texttt{uvotsource} was used to extract the photometry.

\subsection{\mission{NICER} observations\label{subsec:multi_ni}}

The \mission{NICER} data were analysed using \textsc{HEASoft} with the \mission{NICER} data analysis software (version \texttt{10}) and calibration files (version \texttt{20221001}). The \texttt{nicerl2} task is used to generate cleaned X-ray events. Events with overshoot higher than 1.5 or undershoot larger than 300 were removed. The \texttt{nicerl3-spec} task was then used to generate the X-ray spectra for each \mission{NICER} observation. The X-ray spectra were then rebinned to have at least one count per bin. We adopted the \textsc{SCORPEON} model to generate background models for each observation. The \texttt{nicerarf} and \texttt{nicerrmf} tasks were used to generate the response matrix and ancillary file for each observation, respectively. The same procedures were also adopted to re-analyse the \mission{NICER} observations taken during the first two X-ray rising phases (i.e. observation carried out during MJD~59418-59448 and 59600-59641).

\subsection{\mission{Chandra} observations\label{subsec:multi_ch}}

We requested \mission{Chandra} DDT observations of \jsrc, which were performed on 2023 Mar 18, Apr 11, and April 14 with the Advanced CCD Imaging Spectrometer (ACIS). We used the \textsc{CIAO} (\citealt{fruscione_etal06}, version \texttt{4.15}) software package to reduce the \mission{Chandra} data with calibration files CALDB version \texttt{4.10.4}. We reprocessed the \mission{Chandra} data using the \textsc{CIAO} script \texttt{chandra\_repro}. The \textsc{CIAO} task \texttt{dmextract} was used to extract the source and background spectra. We extracted the source spectra using a circular region with a radius of $2''$. The background spectra were extracted using an annulus (concentric with the source) region with an inner and outer radius of $6''$ and $20''$, respectively. The response files were generated using the \texttt{mkacisrmf} and \texttt{mkarf} tasks. The position of the X-ray flare measured from \mission{Chandra} is $(\text{RA, Dec})=(\text{04:56:49.81}, -20\degr37\arcmin47.98\arcsec)$ with a $2\sigma$ uncertainty of $0.54\arcsec$ (Appendix~\ref{sec:ch_pos}), consistent with the centre of the host galaxy.

\subsection{ATCA radio observations\label{subsec:atca}}
We observed the coordinates of \jsrc nine times with the Australia Telescope Compact Array (ATCA) between 2022 Aug and 2023 Mar, in addition to the five observations between 2021 Mar and 2022 Apr reported in \citet{liu_etal2023}. We observed the target during the X-ray outburst phase as this was previously when radio emission had been detected and therefore observed the target with the array in various configurations. In each observation, we used the dual 5.5\,GHz and 9\,GHz receiver, placing the 2x2\,GHz of bandwidth split into 2048x1\,MHz channels at a central frequency of 5.5\,GHz and 9\,GHz. Data were reduced in the Common Astronomy Software Application \citep[CASA, verison \texttt{5.6.3},][]{CASApaper} using standard procedures, including flux and bandpass calibration with PKS 1934-638 and phase calibration with PKS 0454-234. Additionally, we carried out one round of phase-only self-calibration of the target field at both 5.5 and 9\,GHz, with a typical solution interval of 2~minutes, to produce a good quality image due to a bright AGN in the field. Images of the target field were created with the CASA task \texttt{tclean} and where a source was visible at the location of \jsrc, the flux density was extracted with the CASA task \texttt{imfit} by fitting a Gaussian the size of the synthesised beam.

\begin{figure}[!t]
    \centering
    \includegraphics[width=0.5\textwidth]{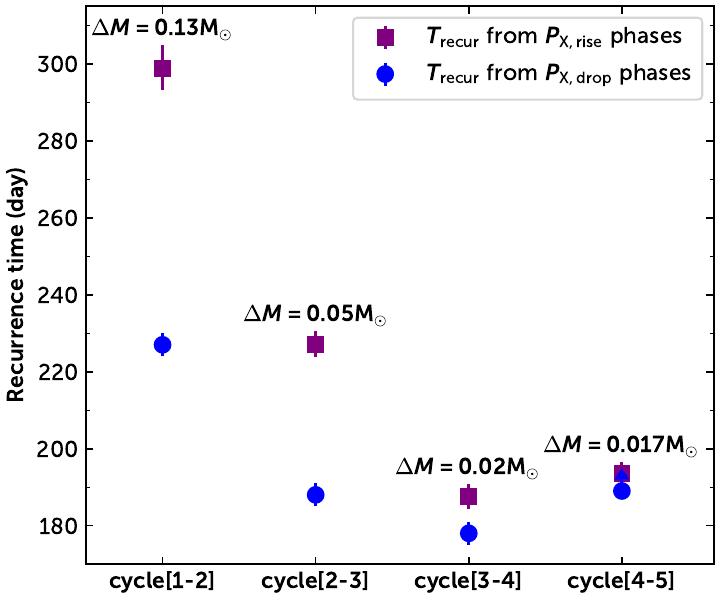}
    \caption{Evolution of the recurrence time. The purple square (blue circle) points represent $\trecur$ measured from the \fxrise (\fxdrop) phases. The error bars mark the $1\sigma$ uncertainties. The upward arrow indicates the $3\sigma$ lower limit. The estimated total mass loss from the star is also marked.\label{fig:evo_of_time}}
\end{figure}

\section{Data analysis and results}\label{sec:results}
\subsection{X-ray spectral modelling}\label{subsec:xray_res}
The \textsc{Xspec} software (version \texttt{12.13.0}, \citealt{arnaud1996}) was used to fit all X-ray spectra using the Cash statistic (\citealt{cash1979}, Cstat in \textsc{Xspec}). As mentioned in \citet{liu_etal2023}, a power-law model (i.e. \texttt{TBabs*zashift*cflux*powerlaw}, hereafter \mpl) is preferred for observations taken at relatively high X-ray flux (i.e. the rest-frame unabsorbed 0.2--2.0~keV flux, \sflux, $\gtrsim5\times10^{-13}\unitflux$). We thus first fitted all the new X-ray spectra with the \mpl model. The Galactic column density is fixed at $3.3\times10^{20}~\mathrm{cm^{-2}}$.

For \mission{NICER} data, we first fitted the total X-ray spectra over the 0.25--10.0\,keV range with the backgrounds model generated using \textsc{SCORPEON}. We then re-fitted the data by adding the \mpl model. A $3\sigma$ upper limit for \sflux was estimated for observations in which the fit did not improve significantly (i.e. $\Delta C_\text{stat}<11.8$) after adding the \mpl model. A strong oxygen K$\alpha$ line is presented in some of the \mission{NICER} data. The \texttt{niscorpv22\_swcxok\_norm} parameter in the background model was thus left free to properly model the oxygen K$\alpha$ line in those observations. We fitted the background subtracted spectra with the \mpl model for the data from the other missions. The \mission{Swift}/XRT X-ray spectra were fitted over the 0.3--5.0\,keV energy range, while the 0.5--5.0\,keV energy band was used for \mission{Chandra} observations. For \mission{XMM-Newton} data, we jointly fitted the data from the three EPIC cameras over the 0.2--5.0\,keV energy range for X5 and X7 (0.2--2.0~keV for X6). The best-fitting values and the 90 percent uncertainties were calculated for the \sflux and photon index parameters. In addition, the 68 percent uncertainties for the \sflux were also estimated for observations taken during the \fxrise phase. The $3\sigma$ upper limits of \sflux were calculated for non-detections using either the \mpl or a disk model (see below). The details of the fitting results are listed in Table\,\ref{tab:logxray}.

\citet{liu_etal2023} noticed that the UV-to-X-ray SEDs of \jsrc can be described by a multi-colour disk model (\texttt{TBabs*zashift*cflux*diskbb}, hereafter \mmcd) when the X-ray flux is low. Thus, the \mmcd model was also used to fitted the X-ray spectra at the early stage of the \fxrise phase, that is, observations taken between MJD~59600 and 59620 in cycle 3 and X6 in cycle 5. \jsrc was detected only on MJD~59619 (\mission{NICER}, ObsID: 4595020126, hereafter N26) and on MJD~59999 (X6). The best-fitting $T_\text{in}$ and \sflux are $193^{+43}_{-25}~(78_{-18}^{+24})~$eV and $2.0^{+0.5}_{-0.3}\times10^{-14}~(2.3_{-0.5}^{+1.2}\times10^{-13})~\unitflux$ for X6 (N26), respectively. The \mpl model resulted in a higher \sflux of $3.3^{+1.0}_{-0.6}\times10^{-14}~(4.6^{+1.1}_{-1.6}\times10^{-13})~\unitflux$ with photon index of $3.3^{+0.3}_{-0.4}~(5.3^{+0.7}_{-0.8})$ for X6 (N26). Both the \mmcd and the \mpl model fitted the X6 and N26 spectra well, though the \mpl model gives a slightly better fit ($C_\text{stat}/\text{d.o.f}=136/144$, compared to 152/144) for X6.

In this work, we used the results from the \mmcd model for X6 and N26. The best-fitting $T_\text{in}$ of N26 is comparable to that obtained from the eRASS2 and \mission{Swift} observations (in the range of $\sim50-100\,$eV, \citealt{liu_etal2023}) at similar \sflux (i.e. $\approx4\times10^{-13}\,\unitflux$), while $T_\text{in}$ is much higher during the X6 observation when \jsrc was in a historically low X-ray flux. This suggests a potential change in $T_\text{in}$ during the \fxrise phase. For this reason, different values of $T_\text{in}$ (listed in Table\,\ref{tab:logxray}) were used to calculate the $3\sigma$ flux upper limits for \mission{NICER} observations before N26 during cycle 3.

\subsection{Modelling the X-ray rising phase}\label{sec:fit_rs}

The \fxrise phases during cycles 2, 3, 4 and 5 were captured by our follow-up observations. We also assumed that \jsrc is in the \fxrise phase of cycle1 during the eRASS2 observations. This is justified by the duration of the \fxdrop phase being much shorter than the \fxrise phase, and by the spectral property of eRASS2 (i.e. best described by the \mmcd model) being similar to that at the early stage of the \fxrise phase.

\begin{table}[]
\caption{Results of the X-ray rising phase modelling and estimations of $\trecur$.}\label{tab:fit_res}
\bgroup
\def\arraystretch{0.6}%
\begin{tabular}{@{\extracolsep{\fill}}ccccc}
\hline\hline
\multirow{2}{*}{Parameters} & \multirow{2}{*}{Cycles[1,2,5]}           & \multirow{2}{*}{All cycles}             & \multirow{2}{*}{$\trecur$}      & \multirow{2}{*}{$T_\text{recur, drop}$}\\[1.5mm]
\multirow{2}{*}{}           & \multirow{2}{*}{MJD}                     & \multirow{2}{*}{MJD}                    & \multirow{2}{*}{days}              & \multirow{2}{*}{days}              \\[2mm]\hline
\multirow{2}{*}{$t_1$   }   & \multirow{2}{*}{$59070_{-8}^{+7}$     }  & \multirow{2}{*}{$59065^{+8}_{-9}$     } &                                    &                                    \\
                            &                                          &                                         & \multirow{2}{*}{$299^{+6}_{-6}$ }  & \multirow{2}{*}{$227\pm3$ }  \\
\multirow{2}{*}{$t_2$   }   & \multirow{2}{*}{$59367_{-5}^{+5}$     }  & \multirow{2}{*}{$59363^{+6}_{-6}$     } &                                    &                                    \\
                            &                                          &                                         & \multirow{2}{*}{$227^{+4}_{-3}$ }  & \multirow{2}{*}{$188\pm3$ }  \\
\multirow{2}{*}{$t_3$   }   & \multirow{2}{*}{---	      }        & \multirow{2}{*}{$59591^{+6}_{-5}$     } &                                    &                                    \\
                            &                                          &                                         & \multirow{2}{*}{$188^{+3}_{-3}$ }  & \multirow{2}{*}{$178\pm3$ }  \\
\multirow{2}{*}{$t_4$   }   & \multirow{2}{*}{---                   }  & \multirow{2}{*}{$59778^{+7}_{-7}$     } &                                    &                                    \\
                            &                                          &                                         & \multirow{2}{*}{$193^{+3}_{-3}$ }  & \multirow{2}{*}{$>189$ }  \\
\multirow{2}{*}{$t_5$   }   & \multirow{2}{*}{$59976_{-4}^{+3}$     }  & \multirow{2}{*}{$59972^{+4}_{-5}$     } &                                    &                                    \\
                            &                                          &                                         & \multirow{2}{*}{}                  & \multirow{2}{*}{}                  \\
\multirow{2}{*}{$\log A$}   & \multirow{2}{*}{$-16.2_{-0.8}^{+0.6}$ }  & \multirow{2}{*}{$-17.2^{+1.0}_{-1.2}$ } &                                    &                                    \\
                            &                                          &                                         & \multirow{2}{*}{}                  & \multirow{2}{*}{}                  \\
\multirow{2}{*}{$\beta$ }   & \multirow{2}{*}{$2.6_{-0.3}^{+0.4}$   }  & \multirow{2}{*}{$3.1^{+0.6}_{-0.5}$   } &                                    &                                    \\
                            &                                          &                                         &                                    &                                    \\
\hline
\end{tabular}
\tablefoot{\textit{Parameters}: name of the parameter in the power-law model; \textit{Cycles[1,2,5]}: results by jointly fitting the \fxrise phases of cycles 1, 2, and 5; \textit{All cycles}: results by jointly fitting the \fxrise of all the five cycles; $\trecur$ and $T_\text{recur, drop}$ are the recurrence time estimated using the \fxrise and \fxdrop phases, respectively.}
\egroup
\end{table}
We jointly fitted the five \fxrise phases with a power law function $f_\text{rs,i}(t)=A*(t-t_i)^{\beta}$, where $i=1,2,3,4,5$. We assumed that the normalisation $A$ and power law index $\beta$ are the same for all the five \fxrise phases. The \texttt{lmfit} package is adopted to fit the data with the least squares method. To take into account upper limits and to estimate the uncertainties of the parameters in the model, we generated $10^5$ realisations of the \sflux in the five \fxrise phases from Gaussian distributions for observations of which \jsrc is detected and uniform distributions with a lower limit of zero (excluded) for non-detections. The best-fitting \sflux and the $1\sigma$ uncertainties obtained from X-ray spectral modelling of each observation are used as the means and standard deviations of the Gaussian distributions, respectively. The $3\sigma$ \sflux upper limits for non-detections are used as the upper limits for the uniform distributions. We obtained $10^5$ values for each parameter in the model by fitting the $10^5$ datasets using the least squares method. We estimated the best-fitting values using the median values for each parameter and estimated the $1\sigma$ confidence intervals using the 16 and 84 percentiles of fitting results. Similarly, the recurrence time $\trecur$ for cycle $i$ and $i+1$, the lower, and the upper intervals of the $1\sigma$ confidence intervals are estimated using the median, the 16, and 84 percentile of a sample calculated using $t_{i+1}-t_{i}$. To test whether the results could be affected by the inclusion of flux upper limits, we applied the same procedures to cycles 1, 2, and 5 only (i.e. cycles without upper limits in the \fxrise phases). We found the results to be consistent within $1\sigma$ uncertainties (see Table\,\ref{tab:fit_res}). The fitting results and the estimated $\trecur$ are listed in Table~\ref{tab:fit_res}. We also calculated the recurrence time using the \fxdrop phase (Appendix~\ref{sec:rec_drop}). The estimated values of $\trecur$ are also listed in Table\,\ref{tab:fit_res}. It is clear from Fig.~\ref{fig:evo_of_time} that the $\trecur$ of the X-ray flares in \jsrc show rapid changes. The mass loss for each cycle marked in Fig.~\ref{fig:evo_of_time} are calculated using the method outlined in Sect.~\ref{sec:massloss}.

\subsection{Estimation of the mass loss}\label{sec:massloss}
\begin{figure}
    \centering
    \includegraphics[width=0.5\textwidth]{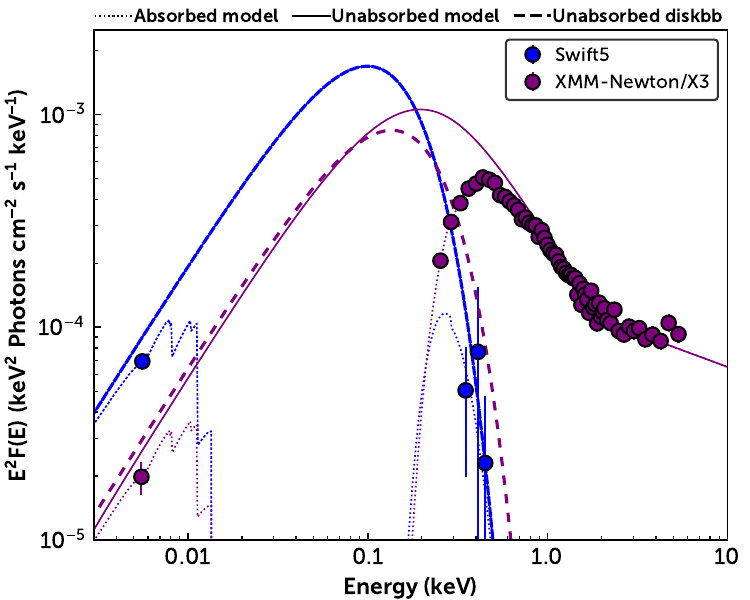}
    \caption{UV to X-ray SEDs at different X-ray flux levels. The solid circles are the unfolded spectra for Swift5 (blue) and \mission{XMM-Newton}/X3 (purple). The dotted lines are the absorbed models. The solid and dashed lines represent the unabsorbed intrinsic model and the diskbb component, respectively.\label{fig:uvxray_sed}}
\end{figure}

Following \citet{liu_etal2023}, a cycle is defined as the time between the start of two consecutive \fxrise phases. The total energy released in each cycle was estimated by
\begin{equation}\label{eq:massloss}
    E_\text{tot}=L_\text{fa} \Delta t_\text{fa}+L_\text{pl} \Delta t_\text{pl}+\int_0^{\Delta t_\text{rs}}L_\text{rs}(t)~\text{d}t\text{,}
\end{equation}
where $L_\text{rs}$ is the bolometric luminosity during the \fxrise phase, which is calculated using the best-fitting power law (see Sect.~\ref{sec:fit_rs}), that is, $L_\text{rs}(t)=\kappa 4\pi D_\text{ld}^2 f_\text{rs}(t)$, where $\kappa=L_\text{bol}/L_\text{X, res 0.2-2.0~keV}$ is the bolometric correction factor and $D_\text{ld}$ is the luminosity distance. $L_\text{pl}$ and $L_\text{fa}$ is the average bolometric luminosity during each of the \fxplat and \fxfaint phases, respectively. $\Delta t_\text{fa}$, $\Delta t_\text{pl}$, and $\Delta t_\text{rs}$ are the duration of the \fxfaint, \fxplat, and \fxrise phases, respectively. The \fxfaint phase in cycle 5 was not covered by our follow-up observations. In this work, we assumed $\Delta t_\text{fa}=70~$days for cycle 5. \citet{liu_etal2023} estimated the $\kappa$ to be in the range of $3-20$ during the \fxrise phase, calculated by modelling the UV to X-ray data using either the \mmcd (when \jsrc is X-ray faint) or Comptonized \mmcd model. The $T_\text{in}$ of the \mmcd model is around $40-60~\text{eV}$ and does not change significantly (see also Fig.~10 in \citealt{liu_etal2023}). Following \citet{liu_etal2023}, we adopted a value of $\kappa=15$ to calculate $L_\text{rs}$ during the \fxrise phases. The high quality X-ray and UV data taken during the \fxplat phase of cycle 2 (the third \mission{XMM-Newton} observation, hereafter X3) can be best-fitted with a multi-color disk (with $T_\text{in}$ around 60~eV) Comptonized by two coronae (\mission{XMM-Newton}/X3 in Fig.~\ref{fig:uvxray_sed}, see also Table~3 in \citealt{liu_etal2023} for the best-fitting results for X3). We estimated a value of $\kappa\sim3$ using this model. Both the X-ray and UV only show mild variability during the \fxplat phases, indicating that $\kappa$ will not change significantly. We thus calculated $L_\text{pl}$ assuming $\kappa=3$ for the \fxplat phases. $L_\text{fa}$ is poorly constrained. We thus estimated $L_\text{fa}$ using the \mmcd model ($T_\text{in}\sim45~\text{eV}$) obtained by modelling the UV and X-ray emission from the first Swift observation in cycle 2 (Swift5 in Fig.~\ref{fig:uvxray_sed}). This model resulted in a bolometric luminosity of $L_\text{bol, sw}=9.0\times10^{43}~\unitlumi$. Considering that the peak UV magnitude in \fxfaint phases is $\lesssim 0.6$~mag brighter than that during the \fxrise phases, we thus conservatively estimated $L_\text{fa}$ by multiplying $L_\text{bol, sw}$ by a factor of 1.8, which leads to a value of $1.6\times10^{44}~\unitlumi$.

The values of the parameters used to calculate $E_\text{tot}$ for each cycle can be found in Table~\ref{tab:massloss}. To calculate the total mass loss in each cycle, we assumed that half of the tidally disrupted debris returns to the SMBH and that the radiation efficiency is 0.1. We estimated a total released energy (mass loss) of 1.12 (0.13), 0.46 (0.05), 0.18 (0.02), 0.15 (0.017), and $\gtrsim0.12\times10^{52}~\text{erg}$ ($\gtrsim0.014~\msun$) for cycles 1, 2, 3, 4, and 5, respectively. A lower limit is given for cycle 5 as the \fxplat phase could be longer then the value quoted in Table~\ref{tab:massloss}.

\begin{table}[]
\bgroup
\def\arraystretch{1.1}
\setlength\tabcolsep{3pt}
\caption{Values of the parameters used to calculate the total release energy ($E_\text{tot}$) and mass loss ($\Delta M$) in each cycle. }
\begin{tabular}{llllll}\hline\hline\label{tab:massloss}
                      & cycle 1 & cycle 2 & cycle 3 & cycle 4 & cycle 5 \\\hline
$t_\text{s}$ (MJD)    & 59065   & 59363   & 59591   & 59778   & 59972   \\
$\Delta t_\text{rs}$ (day)  & 100     & 84      & 51      & 39      & 38      \\
$\Delta t_\text{pl}$ (day)  & 120     & 65      & 57      & 61      & $\gtrsim70$      \\
$\Delta t_\text{fa}$ (day)  & 61      & 71      & 68      & 85      & 70      \\
$L_\text{pl}$ ($10^{44}~\unitlumi$)   & 4.6 & 1.1  & 0.70  & 0.27  & 0.13 \\
$E_\text{tot}$ ($10^{52}~\text{erg}$) & 1.12 & 0.46 & 0.18 & 0.15 & $\gtrsim0.12$ \\
$\Delta M$ ($\msun$)  & 0.13    & 0.05    & 0.02   & 0.017   & $\gtrsim0.014$  \\\hline
\end{tabular}
\tablefoot{$t_\text{s}$ is the start date of the cycle. $\Delta t_\text{rs}$, $\Delta t_\text{pl}$, and $\Delta t_\text{fa}$ are the duration for the \fxrise, \fxplat, and \fxfaint phases, respectively. The estimated values for $E_\text{tot}$ and $\Delta M$ are also listed.}
\egroup
\end{table}

\subsection{Radio variability}
\jsrc was mostly undetected after 2022 Aug, with a detection at 9\,GHz on 2022 Oct 05 and at 5.5\,GHz on 2022 Oct 07 during the \fxplat phase of cycle 4 (see Fig.~\ref{fig:multi_lc}) with a significance of 7 and 9$\sigma$, respectively. In order to improve the sensitivity, we additionally stacked the observations from October 2-7 and detected a faint point source at the coordinates of \jsrc. The radio observations reported here indicate that the transient radio source associated with the X-ray outbursts in 2022 Mar and Apr has also been fading, consistent with the decrease in the peak X-ray flux of each cycle (see Fig.~\ref{fig:multi_lc}). A summary of the ATCA radio observations of \jsrc is given in Table \ref{tab:atca}. The long-term radio light curve is shown in Fig.~\ref{fig:radio_lc}.

\section{Discussion\label{sec:discussion}}
We have detected five repeating X-ray flares in \jsrc using the latest data. In addition, repeating transient radio emission has also been detected in \jsrc. These results provide further evidence that \jsrc is a repeating nuclear transient, and make it one of the most promising repeating \ptde candidates. More importantly, our results also revealed rapid evolution of the recurrence time $\trecur$, measured from the \fxrise phases, of the X-ray flares. $\trecur$ decreased by more than 2~months between cycles 1 and 2. It continued to decrease by roughly $40~$days between cycles 2 and 3. Such a rapid change is likely terminated as suggested by an almost constant $\trecur$ ($\sim190~$days) measured from the latest three cycles. Evidence for evolution of $\trecur$ are also found using the values measured from the \fxdrop phases (Appendix~\ref{sec:rec_drop} and Table~\ref{tab:fit_res}). Though, the changes in $\trecur$ are less dramatic than those measured from the \fxrise phases, which may attributed to the changes in the duration of the other phases (see Table~\ref{tab:massloss}). In this work, $\trecur$ derived from the \fxrise phase were used.

The evolution of $\trecur$ has been reported only in a few repeating \ptde candidates. For instance, \asko showed a decrease in the periods with a period derivative of around $-0.0026$ \citep{payne_etal2022, huang_etal2023}, which is much shorter than that found in \jsrc ($\lesssim-0.2$). The X-ray flares in \hlx initially showed a quasi-periodic $\trecur$ of around 1~yr. $\trecur$ then increased by about 1~month in 2013 \citep{godet_etal2014}, and continued to increase until 2018, after which no X-ray outbursts were detected \citep{webb_etal2023}. Unlike in \hlx, we found no evidence for an increase in $\trecur$ in \jsrc as of now. As suggested by \citet{godet_etal2014}, the evolution of $\trecur$ can put strong constraints on the initial mass, the mass loss, and the orbital parameters of the star.

A repeating \ptde is favoured to explain the long-term multi-wavelength light curve of \jsrc \citep{liu_etal2023}. We thus performed simulations to test if the changes in $\trecur$, $\Delta P$, can be explained by \ptdes. We made a grid of hydrodynamics simulations, using the moving-mesh code {\small AREPO} \citep{Arepo,Arepo2,ArepoHydro}, where we examined the change in the orbital period of remnants produced in \ptdes of main-sequence stars by BHs with a mass of $\mbh=10^{5}\msun$. We considered Solar metallicity main-sequence stars with masses of $M_{\star}=1$, 2, and 3\,$\msun$, and a core Hydrogen mass fraction of 0.01 (terminal-age) and 0.3 (middle-age), evolved using the 1D stellar evolution code {\small MESA} \citep{paxton:13,paxton:15,paxton:19,Paxton+2011}, imported into {\small AREPO} with 0.5M cells\footnote{We confirmed that simulations with different resolutions between 0.25M and 1M cells yield converging $\Delta P$.}. We also considered a wide range of the pericenter distance $r_{\rm p}$, $0.1\lesssim r_{\rm p}/r_{\rm t}\lesssim 1.2$ (tidal radius $r_{\rm t}=(\mbh/M_{\star})^{1/3}R_{\star}$), encompassing scenarios from the full disruption to no mass loss. We varied the pericenter distance for a given $M_{\star}$. Meanwhile, we fixed the orbital period of the original orbit to be 300 days, so the stellar orbit in each simulation has a different eccentricity ($e\gtrsim0.99$). The initial separation between the black hole and the star is $5$\,$r_{\rm t}$. We followed the evolution of the remnant after the first pericenter passage of the original star, using the Helmholtz equation of state \citep{HelmholtzEOS} until the post-disruption orbital parameters do not evolve, which occurred when the separation between the black hole and the remnant is $\gtrsim 5r_{\rm t}$. We verified that the total energy in all simulations was conserved within a fractional error of $\lesssim10^{-5}$. We note that a rapid decrease in $\trecur$ requires a BH mass of around $10^{5}\msun$ in our simulations which ran with a limited range of parameter space. Our simualtions with a higher BH mass of $10^{6}\msun$ cannot reproduce $\trecur$ observed in \jsrc. The required BH mass of $10^{5}\msun$ is much smaller than the value quoted in \citet{liu_etal2023}, which is around $10^{7}\msun$. However, as cautioned in \citet{liu_etal2023}, the values of $\mbh$ measured from the $\mbh-\sigma_\star$ relation and the $\sigma^2_\text{rms}-\mbh$ relation exhibit notable differences and have large uncertainties. Thus a $\mbh$ on the order of $10^{5}\msun$ could still be possible for \jsrc.

\begin{figure}
  \includegraphics[width=\columnwidth]{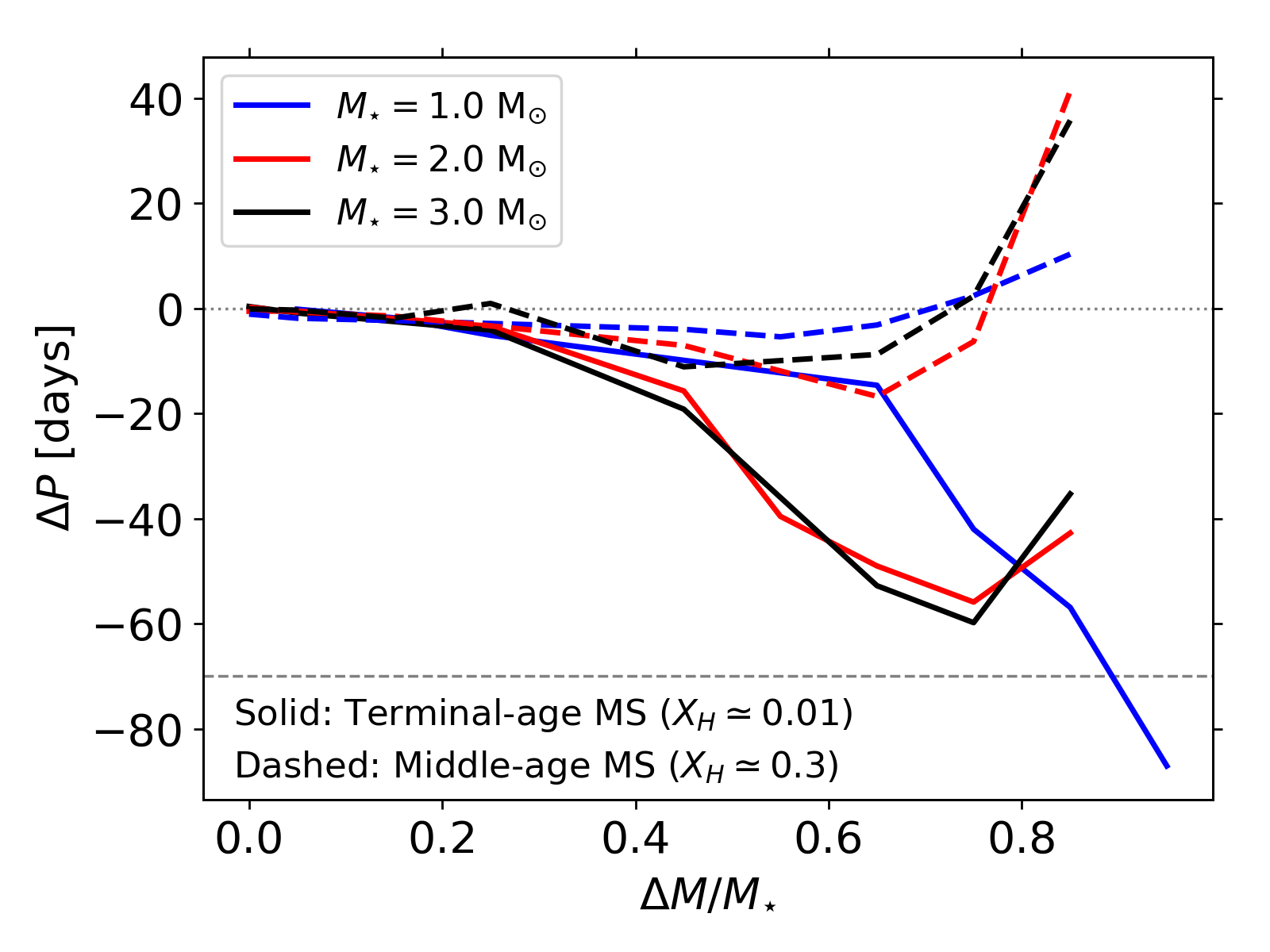}
  \caption{Change in the orbital period $\Delta P$ (in days) for remnants in \ptdes of main-sequence stars with mass of $M_{\star}=\text{1, 2, and 3}~\msun$ in hydrodynamics simulations, as a function of the fractional mass loss $\Delta M/M_{\star}$. The original stars are Solar metallicity main-sequence stars with a core Hydrogen mass fraction of $X_{\rm H}\simeq 0.01$ (Terminal-age, solid lines) and $0.3$ (Middle-age, dashed lines). The simulations suggest that $\Delta P$ in severe \ptdes of terminal-age main-sequence stars can be compatible with the change in $T_{\rm recur}$ between cycles 1 and 2 ($\simeq -70$ days, grey horizontal line) of J0456-20.\label{fig:hydro_sim}}
\end{figure}

Fig.~\ref{fig:hydro_sim} depicts $\Delta P$ as a function of the original stellar mass $M_{\star}$ and the fractional mass loss $\Delta M/M_{\star}$. The general trend is that $\Delta P$ decreases as $\Delta M/M_{\star}$ increases until the fractional mass loss exceeds a critical value $\Delta m_\text{c}$, roughly $\approx0.7-0.8$. Above the critical mass loss, $\Delta P$ starts to increase with $\Delta M/M_{\star}$. This trend was observed for parabolic \ptdes in \citet{ryu_etal2020c}. However, the values of $\Delta P$  strongly depend on the internal structure of the star (i.e. mass and age). For \ptdes of middle-age stars, $\Delta P$ is negative (positive) when $\Delta M/M_{\star}$ is below (above) $\Delta m_\text{c}$, meaning the remnants become more (less) bound than that of the original star before the TDE. For this case, $\Delta P$ is at most $\simeq-10$ days. However, $\Delta P$ is negative across a wider range of $\Delta M/M_{\star}$ and can be as large as $-90$~days for \ptdes of terminal-age stars. Most notably, the large $\Delta P$ (i.e. $\sim-70~$days) observed in \jsrc between cycles 1 and 2 can be reproduced if the original star is a $1\msun$ terminal-age star and loses nearly 80-90\,\% of its mass. While we have not explored the entire parameter space of \ptdes, our simulations suggest that the observed decrease in $\trecur$ may be explained by a severe \ptde of a main-sequence star near the terminal age. The inferred mass loss ($0.8-0.9\,\msun$) is much higher than that estimated for cycle 1 ($\sim0.13\,\msun$, Sect.~\ref{sec:massloss} and Table~\ref{tab:massloss}) of \jsrc. We note that this discrepancy can be alleviated if the radiation efficiency in \jsrc is much lower than the assumed value of 0.1. It is important to also emphasise that our simulation result merely suggests \ptde is a plausible mechanism for generating a transient like \jsrc and does not rule out other potential mechanisms. For instance, \citet{linial_etal2024} summarized several orgins of observed period evolution in repeating nuclear transients. They suggested that the period evolution in \asko is consistent with orbital decay induced by hydrodynamic drag as the star passes through an accretion disk \citep[see also][]{zhou_etal2024}. The repeating flares in \asko could then be powered primarily through dragging-induced stripping of mass from the star. This scenario, however, is unlikely to be the dominant process in \jsrc, as it requires an accretion disk with mass larger than $10~\msun$ (Eq.~20 in \citealt{linial_etal2024}, assuming $\mbh=10^{5}~\msun$) around $r_\text{p}$ to explain the observed changes in $\trecur$ in \jsrc.

\section{Summary\label{sec:summary}}
In this letter, we analysed new multi-wavelength observations for the promising repeating \ptde candidate \jsrc. We have detected five repeating X-ray flares and repeating transient radio emission from \jsrc, providing additional strong evidence that \jsrc is a repeating nuclear transient. In addition, the latest X-ray data also reveals changes in the recurrence time $\trecur$ of the X-ray flare. $\trecur$ initially decreased rapidly from $\sim300~$days to $\sim230$\,days, it continues to decrease by roughly $40~$days per cycle with an indication of constant values of $\sim190$\,days in the latest cycles. Our hydrodynamic simulations show that the large decrease in $\trecur$ can be explained in the \ptde scenario, providing that the original star is a $1\,\msun$ terminal-age star with an initial fractional mass loss of around 80-90\%. Our results suggest that precise estimations of $\trecur$ in repeating \ptdes can provide additional constraints on the initial mass of the disrupted star and mass loss during each passage. They also indicate that repeating \ptdes could be effective tools to probe the stellar/gas dynamics around SMBHs beyond our own Galaxy.

\begin{acknowledgements}
ZL is grateful to the \mission{XMM-Newton}, \mission{Swift}, and \mission{NICER} teams for approving the ToO/DDT requests and arranging the follow-up observations. The hydrodynamics simulations were conducted using computational resources (and/or scientific computing services) at the Max-Planck Computing \& Data Facility. DH acknowledges support from DLR grant FKZ 50 OR 2003. MK is supported by DFG grant KR 3338/4-1 and DLR grant 50 OR 2307. This work was supported by the Australian government through the Australian Research Council’s Discovery Projects funding scheme (DP200102471).

\end{acknowledgements}

\bibliographystyle{aa}
\bibliography{references}

\begin{thebibliography}{44}
\expandafter\ifx\csname natexlab\endcsname\relax\def\natexlab#1{#1}\fi

\bibitem[{{Arcodia} {et~al.}(2021){Arcodia}, {Merloni}, {Nandra}, {Buchner},
  {Salvato}, {Pasham}, {Remillard}, {Comparat}, {Lamer}, {Ponti}, {Malyali},
  {Wolf}, {Arzoumanian}, {Bogensberger}, {Buckley}, {Gendreau}, {Gromadzki},
  {Kara}, {Krumpe}, {Markwardt}, {Ramos-Ceja}, {Rau}, {Schramm}, \&
  {Schwope}}]{arcodia_etal2021}
{Arcodia}, R., {Merloni}, A., {Nandra}, K., {et~al.} 2021, Nature, 592, 704

\bibitem[{Arnaud(1996)}]{arnaud1996}
Arnaud, K.~A. 1996, ASP Conf. Ser., 101, 17

\bibitem[{{Bortolas} {et~al.}(2023){Bortolas}, {Ryu}, {Broggi}, \&
  {Sesana}}]{Bortolas+2023}
{Bortolas}, E., {Ryu}, T., {Broggi}, L., \& {Sesana}, A. 2023, \mnras, 524,
  3026

\bibitem[{{CASA Team} {et~al.}(2022){CASA Team}, {Bean}, {Bhatnagar}, {Castro},
  {Donovan Meyer}, {Emonts}, {Garcia}, {Garwood}, {Golap}, {Gonzalez Villalba},
  {Harris}, {Hayashi}, {Hoskins}, {Hsieh}, {Jagannathan}, {Kawasaki},
  {Keimpema}, {Kettenis}, {Lopez}, {Marvil}, {Masters}, {McNichols},
  {Mehringer}, {Miel}, {Moellenbrock}, {Montesino}, {Nakazato}, {Ott}, {Petry},
  {Pokorny}, {Raba}, {Rau}, {Schiebel}, {Schweighart}, {Sekhar}, {Shimada},
  {Small}, {Steeb}, {Sugimoto}, {Suoranta}, {Tsutsumi}, {van Bemmel},
  {Verkouter}, {Wells}, {Xiong}, {Szomoru}, {Griffith}, {Glendenning}, \&
  {Kern}}]{CASApaper}
{CASA Team}, {Bean}, B., {Bhatnagar}, S., {et~al.} 2022, \pasp, 134, 114501

\bibitem[{{Cash}(1979)}]{cash1979}
{Cash}, W. 1979, \apj, 228, 939

\bibitem[{{Cufari} {et~al.}(2022){Cufari}, {Coughlin}, \&
  {Nixon}}]{cufari_etal2022}
{Cufari}, M., {Coughlin}, E.~R., \& {Nixon}, C.~J. 2022, \apjl, 929, L20

\bibitem[{{Cufari} {et~al.}(2023){Cufari}, {Nixon}, \&
  {Coughlin}}]{cufari_etal2023}
{Cufari}, M., {Nixon}, C.~J., \& {Coughlin}, E.~R. 2023, \mnras, 520, L38

\bibitem[{{Davis} {et~al.}(2011){Davis}, {Narayan}, {Zhu}, {Barret}, {Farrell},
  {Godet}, {Servillat}, \& {Webb}}]{davis_etal2011}
{Davis}, S.~W., {Narayan}, R., {Zhu}, Y., {et~al.} 2011, \apj, 734, 111

\bibitem[{{Evans} {et~al.}(2009){Evans}, {Beardmore}, {Page}, {Osborne},
  {O'Brien}, {Willingale}, {Starling}, {Burrows}, {Godet}, {Vetere}, {Racusin},
  {Goad}, {Wiersema}, {Angelini}, {Capalbi}, {Chincarini}, {Gehrels}, {Kennea},
  {Margutti}, {Morris}, {Mountford}, {Pagani}, {Perri}, {Romano}, \&
  {Tanvir}}]{evans_etal2009}
{Evans}, P.~A., {Beardmore}, A.~P., {Page}, K.~L., {et~al.} 2009, \mnras, 397,
  1177

\bibitem[{{Evans} {et~al.}(2023){Evans}, {Nixon}, {Campana},
  {Charalampopoulos}, {Perley}, {Breeveld}, {Page}, {Oates}, {Eyles-Ferris},
  {Malesani}, {Izzo}, {Goad}, {O'Brien}, {Osborne}, \&
  {Sbarufatti}}]{evans_etal2023}
{Evans}, P.~A., {Nixon}, C.~J., {Campana}, S., {et~al.} 2023, Nature Astronomy,
  7, 1368

\bibitem[{{Fruscione} {et~al.}(2006){Fruscione}, {McDowell}, {Allen},
  {Brickhouse}, {Burke}, {Davis}, {Durham}, {Elvis}, {Galle}, {Harris},
  {Huenemoerder}, {Houck}, {Ishibashi}, {Karovska}, {Nicastro}, {Noble},
  {Nowak}, {Primini}, {Siemiginowska}, {Smith}, \& {Wise}}]{fruscione_etal06}
{Fruscione}, A., {McDowell}, J.~C., {Allen}, G.~E., {et~al.} 2006, in Society
  of Photo-Optical Instrumentation Engineers (SPIE) Conference Series, Vol.
  6270, 62701V

\bibitem[{{Gabriel} {et~al.}(2004){Gabriel}, {Denby}, {Fyfe}, {Hoar}, {Ibarra},
  {Ojero}, {Osborne}, {Saxton}, {Lammers}, \& {Vacanti}}]{gabriel_etal2004}
{Gabriel}, C., {Denby}, M., {Fyfe}, D.~J., {et~al.} 2004, in Astronomical
  Society of the Pacific Conference Series, Vol. 314, Astronomical Data
  Analysis Software and Systems (ADASS) XIII, ed. F.~{Ochsenbein}, M.~G.
  {Allen}, \& D.~{Egret}, 759

\bibitem[{{Giustini} {et~al.}(2020){Giustini}, {Miniutti}, \&
  {Saxton}}]{giustini_etal2020}
{Giustini}, M., {Miniutti}, G., \& {Saxton}, R.~D. 2020, A\&A, 636, L2

\bibitem[{{Godet} {et~al.}(2014){Godet}, {Lombardi}, {Antonini}, {Barret},
  {Webb}, {Vingless}, \& {Thomas}}]{godet_etal2014}
{Godet}, O., {Lombardi}, J.~C., {Antonini}, F., {et~al.} 2014, \apj, 793, 105

\bibitem[{{Guillochon} \& {Ramirez-Ruiz}(2013)}]{guillochon_etal2013}
{Guillochon}, J. \& {Ramirez-Ruiz}, E. 2013, \apj, 767, 25

\bibitem[{{Guolo} {et~al.}(2024){Guolo}, {Pasham}, {Zaja{\v{c}}ek}, {Coughlin},
  {Gezari}, {Sukov{\'a}}, {Wevers}, {Witzany}, {Tombesi}, {van Velzen},
  {Alexander}, {Yao}, {Arcodia}, {Karas}, {Miller-Jones}, {Remillard},
  {Gendreau}, \& {Ferrara}}]{guolo_etal2023}
{Guolo}, M., {Pasham}, D.~R., {Zaja{\v{c}}ek}, M., {et~al.} 2024, Nature
  Astronomy [\eprint[arXiv]{2309.03011}]

\bibitem[{{Hayasaki} {et~al.}(2013){Hayasaki}, {Stone}, \&
  {Loeb}}]{hayasaki_etal2013}
{Hayasaki}, K., {Stone}, N., \& {Loeb}, A. 2013, MNRAS, 434, 909

\bibitem[{{Huang} {et~al.}(2023){Huang}, {Jiang}, {Shen}, {Wang}, \&
  {Sheng}}]{huang_etal2023}
{Huang}, S., {Jiang}, N., {Shen}, R.-F., {Wang}, T., \& {Sheng}, Z. 2023,
  \apjl, 956, L46

\bibitem[{{Krolik} {et~al.}(2020){Krolik}, {Piran}, \& {Ryu}}]{Krolikc+2020}
{Krolik}, J., {Piran}, T., \& {Ryu}, T. 2020, \apj, 904, 68

\bibitem[{{Linial} \& {Quataert}(2024)}]{linial_etal2024}
{Linial}, I. \& {Quataert}, E. 2024, \mnras, 527, 4317

\bibitem[{{Liu} {et~al.}(2023){Liu}, {Malyali}, {Krumpe}, {Homan}, {Goodwin},
  {Grotova}, {Kawka}, {Rau}, {Merloni}, {Anderson}, {Miller-Jones},
  {Markowitz}, {Ciroi}, {Di Mille}, {Schramm}, {Tang}, {Buckley}, {Gromadzki},
  {Jin}, \& {Buchner}}]{liu_etal2023}
{Liu}, Z., {Malyali}, A., {Krumpe}, M., {et~al.} 2023, \aap, 669, A75

\bibitem[{{Malyali} {et~al.}(2023){Malyali}, {Liu}, {Rau}, {Grotova},
  {Merloni}, {Goodwin}, {Anderson}, {Miller-Jones}, {Kawka}, {Arcodia},
  {Buchner}, {Nandra}, {Homan}, \& {Krumpe}}]{malyali_etal2023}
{Malyali}, A., {Liu}, Z., {Rau}, A., {et~al.} 2023, \mnras, 520, 3549

\bibitem[{{Melchor} {et~al.}(2024){Melchor}, {Mockler}, {Naoz}, {Rose}, \&
  {Ramirez-Ruiz}}]{melchor_etal2024}
{Melchor}, D., {Mockler}, B., {Naoz}, S., {Rose}, S.~C., \& {Ramirez-Ruiz}, E.
  2024, \apj, 960, 39

\bibitem[{{Miniutti} {et~al.}(2023){Miniutti}, {Giustini}, {Arcodia}, {Saxton},
  {Read}, {Bianchi}, \& {Alexander}}]{miniutti_etal2023}
{Miniutti}, G., {Giustini}, M., {Arcodia}, R., {et~al.} 2023, A\&A, 670, A93

\bibitem[{{Miniutti} {et~al.}(2019){Miniutti}, {Saxton}, {Giustini},
  {Alexander}, {Fender}, {Heywood}, {Monageng}, {Coriat}, {Tzioumis}, {Read},
  {Knigge}, {Gandhi}, {Pretorius}, \&
  {Ag{\'\i}s-Gonz{\'a}lez}}]{miniutti_etal2019}
{Miniutti}, G., {Saxton}, R.~D., {Giustini}, M., {et~al.} 2019, Nature, 573,
  381

\bibitem[{{Nixon} \& {Coughlin}(2022)}]{nixon_etal2022}
{Nixon}, C.~J. \& {Coughlin}, E.~R. 2022, \apjl, 927, L25

\bibitem[{{Pakmor} {et~al.}(2016){Pakmor}, {Springel}, {Bauer}, {Mocz},
  {Munoz}, {Ohlmann}, {Schaal}, \& {Zhu}}]{ArepoHydro}
{Pakmor}, R., {Springel}, V., {Bauer}, A., {et~al.} 2016, \mnras, 455, 1134

\bibitem[{{Paxton} {et~al.}(2011){Paxton}, {Bildsten}, {Dotter}, {Herwig},
  {Lesaffre}, \& {Timmes}}]{Paxton+2011}
{Paxton}, B., {Bildsten}, L., {Dotter}, A., {et~al.} 2011, \apjs, 192, 3

\bibitem[{{Paxton} {et~al.}(2013){Paxton}, {Cantiello}, {Arras}, {Bildsten},
  {Brown}, {Dotter}, {Mankovich}, {Montgomery}, {Stello}, {Timmes}, \&
  {Townsend}}]{paxton:13}
{Paxton}, B., {Cantiello}, M., {Arras}, P., {et~al.} 2013, \apjs, 208, 4

\bibitem[{{Paxton} {et~al.}(2015){Paxton}, {Marchant}, {Schwab}, {Bauer},
  {Bildsten}, {Cantiello}, {Dessart}, {Farmer}, {Hu}, {Langer}, {Townsend},
  {Townsley}, \& {Timmes}}]{paxton:15}
{Paxton}, B., {Marchant}, P., {Schwab}, J., {et~al.} 2015, \apjs, 220, 15

\bibitem[{{Paxton} {et~al.}(2019){Paxton}, {Smolec}, {Schwab}, {Gautschy},
  {Bildsten}, {Cantiello}, {Dotter}, {Farmer}, {Goldberg}, {Jermyn}, {Kanbur},
  {Marchant}, {Thoul}, {Townsend}, {Wolf}, {Zhang}, \& {Timmes}}]{paxton:19}
{Paxton}, B., {Smolec}, R., {Schwab}, J., {et~al.} 2019, \apjs, 243, 10

\bibitem[{{Payne} {et~al.}(2022){Payne}, {Shappee}, {Hinkle}, {Holoien},
  {Auchettl}, {Kochanek}, {Stanek}, {Thompson}, {Tucker}, {Armstrong}, {Boyd},
  {Brimacombe}, {Cornect}, {Huber}, {Jha}, \& {Lin}}]{payne_etal2022}
{Payne}, A.~V., {Shappee}, B.~J., {Hinkle}, J.~T., {et~al.} 2022, ApJ, 926, 142

\bibitem[{{Payne} {et~al.}(2021){Payne}, {Shappee}, {Hinkle}, {Vallely},
  {Kochanek}, {Holoien}, {Auchettl}, {Stanek}, {Thompson}, {Neustadt},
  {Tucker}, {Armstrong}, {Brimacombe}, {Cacella}, {Cornect}, {Denneau},
  {Fausnaugh}, {Flewelling}, {Grupe}, {Heinze}, {Lopez}, {Monard}, {Prieto},
  {Schneider}, {Sheppard}, {Tonry}, \& {Weiland}}]{payne_etal2021}
{Payne}, A.~V., {Shappee}, B.~J., {Hinkle}, J.~T., {et~al.} 2021, ApJ, 910, 125

\bibitem[{{Planck Collaboration} {et~al.}(2020){Planck Collaboration},
  {Aghanim}, {Akrami}, {Ashdown}, {Aumont}, {Baccigalupi}, {Ballardini},
  {Banday}, {Barreiro}, {Bartolo}, {Basak}, {Battye}, {Benabed}, {Bernard},
  {Bersanelli}, {Bielewicz}, {Bock}, {Bond}, {Borrill}, {Bouchet}, {Boulanger},
  {Bucher}, {Burigana}, {Butler}, {Calabrese}, {Cardoso}, {Carron},
  {Challinor}, {Chiang}, {Chluba}, {Colombo}, {Combet}, {Contreras}, {Crill},
  {Cuttaia}, {de Bernardis}, {de Zotti}, {Delabrouille}, {Delouis}, {Di
  Valentino}, {Diego}, {Dor{\'e}}, {Douspis}, {Ducout}, {Dupac}, {Dusini},
  {Efstathiou}, {Elsner}, {En{\ss}lin}, {Eriksen}, {Fantaye}, {Farhang},
  {Fergusson}, {Fernandez-Cobos}, {Finelli}, {Forastieri}, {Frailis},
  {Fraisse}, {Franceschi}, {Frolov}, {Galeotta}, {Galli}, {Ganga},
  {G{\'e}nova-Santos}, {Gerbino}, {Ghosh}, {Gonz{\'a}lez-Nuevo}, {G{\'o}rski},
  {Gratton}, {Gruppuso}, {Gudmundsson}, {Hamann}, {Handley}, {Hansen},
  {Herranz}, {Hildebrandt}, {Hivon}, {Huang}, {Jaffe}, {Jones}, {Karakci},
  {Keih{\"a}nen}, {Keskitalo}, {Kiiveri}, {Kim}, {Kisner}, {Knox},
  {Krachmalnicoff}, {Kunz}, {Kurki-Suonio}, {Lagache}, {Lamarre}, {Lasenby},
  {Lattanzi}, {Lawrence}, {Le Jeune}, {Lemos}, {Lesgourgues}, {Levrier},
  {Lewis}, {Liguori}, {Lilje}, {Lilley}, {Lindholm}, {L{\'o}pez-Caniego},
  {Lubin}, {Ma}, {Mac{\'\i}as-P{\'e}rez}, {Maggio}, {Maino}, {Mandolesi},
  {Mangilli}, {Marcos-Caballero}, {Maris}, {Martin}, {Martinelli},
  {Mart{\'\i}nez-Gonz{\'a}lez}, {Matarrese}, {Mauri}, {McEwen}, {Meinhold},
  {Melchiorri}, {Mennella}, {Migliaccio}, {Millea}, {Mitra},
  {Miville-Desch{\^e}nes}, {Molinari}, {Montier}, {Morgante}, {Moss}, {Natoli},
  {N{\o}rgaard-Nielsen}, {Pagano}, {Paoletti}, {Partridge}, {Patanchon},
  {Peiris}, {Perrotta}, {Pettorino}, {Piacentini}, {Polastri}, {Polenta},
  {Puget}, {Rachen}, {Reinecke}, {Remazeilles}, {Renzi}, {Rocha}, {Rosset},
  {Roudier}, {Rubi{\~n}o-Mart{\'\i}n}, {Ruiz-Granados}, {Salvati}, {Sandri},
  {Savelainen}, {Scott}, {Shellard}, {Sirignano}, {Sirri}, {Spencer},
  {Sunyaev}, {Suur-Uski}, {Tauber}, {Tavagnacco}, {Tenti}, {Toffolatti},
  {Tomasi}, {Trombetti}, {Valenziano}, {Valiviita}, {Van Tent}, {Vibert},
  {Vielva}, {Villa}, {Vittorio}, {Wandelt}, {Wehus}, {White}, {White},
  {Zacchei}, \& {Zonca}}]{planck_etal2020}
{Planck Collaboration}, {Aghanim}, N., {Akrami}, Y., {et~al.} 2020, \aap, 641,
  A6

\bibitem[{{Ryu} {et~al.}(2020){Ryu}, {Krolik}, {Piran}, \&
  {Noble}}]{ryu_etal2020c}
{Ryu}, T., {Krolik}, J., {Piran}, T., \& {Noble}, S.~C. 2020, ApJ, 904, 100

\bibitem[{{Springel}(2010)}]{Arepo}
{Springel}, V. 2010, \mnras, 401, 791

\bibitem[{{Str{\"u}der} {et~al.}(2001){Str{\"u}der}, {Briel}, {Dennerl},
  {Hartmann}, {Kendziorra}, {Meidinger}, {Pfeffermann}, {Reppin}, {Aschenbach},
  {Bornemann}, {Br{\"a}uninger}, {Burkert}, {Elender}, {Freyberg}, {Haberl},
  {Hartner}, {Heuschmann}, {Hippmann}, {Kastelic}, {Kemmer}, {Kettenring},
  {Kink}, {Krause}, {M{\"u}ller}, {Oppitz}, {Pietsch}, {Popp}, {Predehl},
  {Read}, {Stephan}, {St{\"o}tter}, {Tr{\"u}mper}, {Holl}, {Kemmer}, {Soltau},
  {St{\"o}tter}, {Weber}, {Weichert}, {von Zanthier}, {Carathanassis}, {Lutz},
  {Richter}, {Solc}, {B{\"o}ttcher}, {Kuster}, {Staubert}, {Abbey}, {Holland},
  {Turner}, {Balasini}, {Bignami}, {La Palombara}, {Villa}, {Buttler},
  {Gianini}, {Lain{\'e}}, {Lumb}, \& {Dhez}}]{struder_etal2001}
{Str{\"u}der}, L., {Briel}, U., {Dennerl}, K., {et~al.} 2001, \aap, 365, L18

\bibitem[{{Timmes} \& {Swesty}(2000)}]{HelmholtzEOS}
{Timmes}, F.~X. \& {Swesty}, F.~D. 2000, \apjs, 126, 501

\bibitem[{{Turner} {et~al.}(2001){Turner}, {Abbey}, {Arnaud}, {Balasini},
  {Barbera}, {Belsole}, {Bennie}, {Bernard}, {Bignami}, {Boer}, {Briel},
  {Butler}, {Cara}, {Chabaud}, {Cole}, {Collura}, {Conte}, {Cros}, {Denby},
  {Dhez}, {Di Coco}, {Dowson}, {Ferrando}, {Ghizzardi}, {Gianotti}, {Goodall},
  {Gretton}, {Griffiths}, {Hainaut}, {Hochedez}, {Holland}, {Jourdain},
  {Kendziorra}, {Lagostina}, {Laine}, {La Palombara}, {Lortholary}, {Lumb},
  {Marty}, {Molendi}, {Pigot}, {Poindron}, {Pounds}, {Reeves}, {Reppin},
  {Rothenflug}, {Salvetat}, {Sauvageot}, {Schmitt}, {Sembay}, {Short},
  {Spragg}, {Stephen}, {Str{\"u}der}, {Tiengo}, {Trifoglio}, {Tr{\"u}mper},
  {Vercellone}, {Vigroux}, {Villa}, {Ward}, {Whitehead}, \&
  {Zonca}}]{turner_etal2001}
{Turner}, M.~J.~L., {Abbey}, A., {Arnaud}, M., {et~al.} 2001, \aap, 365, L27

\bibitem[{{Webb} {et~al.}(2012){Webb}, {Cseh}, {Lenc}, {Godet}, {Barret},
  {Corbel}, {Farrell}, {Fender}, {Gehrels}, \& {Heywood}}]{webb_etal2012}
{Webb}, N., {Cseh}, D., {Lenc}, E., {et~al.} 2012, Science, 337, 554

\bibitem[{{Webb} {et~al.}(2023){Webb}, {Barret}, {Godet}, {Gupta}, {Lin},
  {Quintin}, \& {Tranin}}]{webb_etal2023}
{Webb}, N.~A., {Barret}, D., {Godet}, O., {et~al.} 2023, Astronomische
  Nachrichten, 344, easna.20230051

\bibitem[{{Weinberger} {et~al.}(2020){Weinberger}, {Springel}, \&
  {Pakmor}}]{Arepo2}
{Weinberger}, R., {Springel}, V., \& {Pakmor}, R. 2020, \apjs, 248, 32

\bibitem[{{Wevers} {et~al.}(2023){Wevers}, {Coughlin}, {Pasham}, {Guolo},
  {Sun}, {Wen}, {Jonker}, {Zabludoff}, {Malyali}, {Arcodia}, {Liu}, {Merloni},
  {Rau}, {Grotova}, {Short}, \& {Cao}}]{wevers_etal2023}
{Wevers}, T., {Coughlin}, E.~R., {Pasham}, D.~R., {et~al.} 2023, ApJL, 942, L33

\bibitem[{{Zhou} {et~al.}(2024){Zhou}, {Huang}, {Guo}, {Li}, \&
  {Pan}}]{zhou_etal2024}
{Zhou}, C., {Huang}, L., {Guo}, K., {Li}, Y.-P., \& {Pan}, Z. 2024, arXiv
  e-prints, arXiv:2401.11190

\end{thebibliography}

\begin{appendix}
\section{\mission{Chandra} astrometric correction}\label{sec:ch_pos}
\begin{figure}
  \includegraphics[width=\columnwidth]{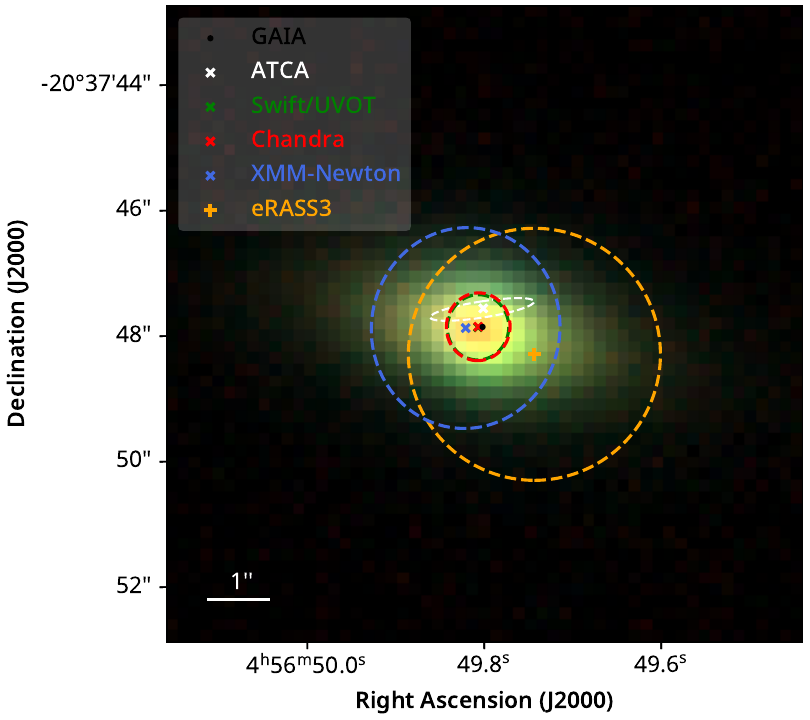}
  \caption{Updated multi-wavelength positions of \jsrc. The black dot marks the position given in the \mission{Gaia} EDR3. The crosses represent the positions measured from different instruments, and the circles indicate the $2\sigma$ positional uncertainties: \mission{Swift}/UVOT (green, $2\sigma=0.52\arcsec$), \mission{Chandra} (red, $2\sigma=0.54\arcsec$), ATCA (white, $2\sigma=1.0\arcsec$), \mission{XMM-Newton} (royalblue, $2\sigma=1.6\arcsec$), and eRASS3 (orange, $2\sigma=2.0\arcsec$).\label{fig:fc_img}}
\end{figure}

We ran the \textsc{wavdetect} tool on the \mission{Chandra} data to generate a source list for each observation. \jsrc was detected in all the three \mission{Chandra} observations. We used the source positions provided in the Legacy Survey DR10 catalogue as a reference. The overall 90 per cent absolute astrometry uncertainty of \mission{Chandra} is $\sim1.11$\,arcsec\footnote{\url{http://cxc.cfa.harvard.edu/cal/ASPECT/celmon}}. We thus cross-matched the positions of the X-ray source measured from the \mission{Chandra} observations with the source positions listed in the DR10 catalogue using a radius of 1.1\,arcsec. We selected sources within 3\,arcmin of the aimpoint of the telescope. This resulted in three sources (excluding \jsrc, see \ref{tab:as_corr}) that are detected in both X-ray and DR10 in the second \mission{Chandra} observation (only two for the other two observations). The \textsc{CIAO} task \textsc{wcs\_match} and \textsc{wcs\_update} were then used to correct and update the aspect ratio. The residual rms scatter in the corrected X-ray positions of the LS10 sources is 0.28\,arcsec, which corresponds to a $2\sigma$ position error of $\approx0.54$\,arcsec (assuming Rayleigh distribution). The astrometric corrected \mission{Chandra} position is consistent with the centre of the host galaxy (see Fig.~\ref{fig:fc_img}).

\begin{table}[]
\bgroup
\def\arraystretch{1.1}
\caption{Astrometric correction for \mission{Chandra} observation.\label{tab:as_corr}}
\begin{tabular}{llc}
\hline \hline
 & Object & Coordinates/Seperations \\\hline
\multirow{4}{*}{LS DR10}              & \jsrc & 04:56:49.80, -20\degr37\arcmin47.99\arcsec \\
                                      & Ref1  & 04:56:57.39, -20\degr35\arcmin12.75\arcsec \\
                                      & Ref2  & 04:56:58.64, -20\degr35\arcmin57.25\arcsec \\
                                      & Ref3  & 04:56:58.70, -20\degr37\arcmin47.65\arcsec \\
\multirow[b]{2}{*}{\mission{Chandra}} & \jsrc & 04:56:49.81, -20\degr37\arcmin47.98\arcsec \\
                                      & Ref1  & 04:56:57.39, -20\degr35\arcmin13.25\arcsec \\
\multirow[t]{2}{*}{With correction}   & Ref2  & 04:56:58.66, -20\degr35\arcmin57.49\arcsec \\
                                      & Ref3  & 04:56:58.73, -20\degr37\arcmin48.58\arcsec \\\hline
\multirow[b]{2}{*}{\mission{Chandra}} & \jsrc & 04:56:49.82, -20\degr37\arcmin48.54\arcsec \\
                                      & Ref1  & 04:56:57.38, -20\degr35\arcmin12.69\arcsec \\
\multirow[t]{3}{*}{No correction}     & Ref2  & 04:56:58.64, -20\degr35\arcmin56.94\arcsec \\
                                      & Ref3  & 04:56:58.71, -20\degr37\arcmin48.03\arcsec \\\hline
\multirow{4}{*}{Seperation}           & \jsrc & 0.07\arcsec  \\
                                      & Ref1  & 0.18\arcsec  \\
                                      & Ref2  & 0.32\arcsec  \\
                                      & Ref3  & 0.41\arcsec \\\hline
\end{tabular}
\tablefoot{Ref1--3 are the reference sources used for performing astrometric correction for \mission{Chandra}. The Separations are calculated using the astrometric corrected \mission{Chandra} positions and the LS DR10 positions.}
\egroup
\end{table}

\section{Estimation of the recurrence time using the X-ray drop phase}\label{sec:rec_drop}
The recurrence time can also be determined using the \fxdrop phases. The start dates of the \fxdrop phases, $t_\text{s, drop}$, can be well constrained in a model-independent way due to their short duration (i.e. \sflux drops by a factor of $\gtrsim10$ within one week). We estimated $t_\text{s, drop}$ in the range of MJD~$[59285, 59291]$, $[59512, 59518]$, and $[59699, 59706]$, and $[59878, 59883]$ for cycles 1, 2, 3, and 4, respectively. The \fxdrop phase was not observed in cycle 5 because \jsrc was blocked by the Sun. However, a lower limit of $>\text{MJD}~60072$ can be given. We generated $10^5$ realisations of $t_\text{s, drop}$ for each cycle in cycles 1--4, assuming uniform distributions of $t_\text{s, drop}$ in the estimated MJD ranges. $\trecur$ and the $1\sigma$ intervals can be estimated using the median and the 16 and the 84 percentiles of the difference between consecutive $t_\text{s, drop}$ for cycles 1--4. We calculated a $3\sigma$ upper limit of $189~$days for $\trecur$ between cycles 4 and 5. The estimated $\trecur$ are listed in Table\,\ref{tab:fit_res}.

\section{Details of the radio and X-ray observations}\label{sec:obs_log}
\begin{figure}
    \centering
    \includegraphics[width=\columnwidth]{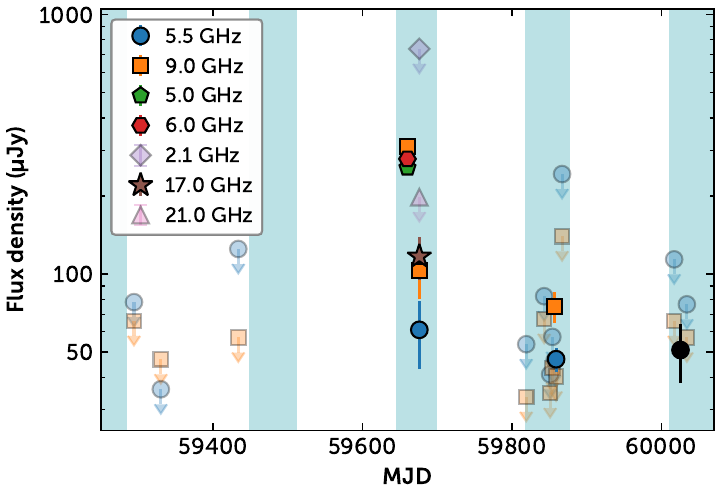}
    \caption{Evolution of the radio flux density. The error bars mark the $1\sigma$ uncertainties. The downward arrows represent the $3\sigma$ upper limits. The black circle shows the flux density at $5.5~$GHz measured from the stacked observation from 2023 Mar. The light cyan regions indicate the \fxplat phase in each cycle.}
    \label{fig:radio_lc}
\end{figure}

Table~\ref{tab:atca} lists the ATCA radio observations for \jsrc. \jsrc was detected only in the \fxplat phase during cycles 3 and 4. The long-term radio variability is shown in Fig.~\ref{fig:radio_lc}.

\begin{table}
\caption{ATCA radio observations of \jsrc}
\centering
\label{tab:atca}
\def\arraystretch{1.0}%
\setlength\tabcolsep{2.7ex}
\begin{tabular}{@{\extracolsep{\fill}}lccc}
\hline\hline
Date  & Frequency  & Array   & Flux Density  \\
      & (GHz) & config.  & ($\mu$Jy)\\\hline
\multirow[c]{2}{*}{2021-03-21} & 5.5 & \multirow[c]{2}{*}{6D}     & $<78$ \\
           & 9   &        & $<66$ \\\hline
\multirow{2}{*}{2021-04-26} & 5.5 & \multirow[c]{2}{*}{6D}     & $<36$ \\
           & 9   &        & $<47$ \\\hline
\multirow{2}{*}{2021-08-07} & 5.5 & \multirow[c]{2}{*}{EW367}  & $<125$ \\
           & 9   &        & $<57$ \\\hline
\multirow{3}{*}{2022-03-22} & 5   & \multirow[c]{3}{*}{6A}     & $257\pm11$ \\
           & 6   &        & $278\pm11$ \\
           & 9   &        & $311\pm12$ \\\hline
\multirow{5}{*}{2022-04-06} & 2.1 & \multirow[c]{5}{*}{6A}     &  $<738$ \\
           & 5.5 &        & $61\pm18$ \\
           & 9   &        & $103\pm23$ \\
           & 17  &        & $117\pm22$ \\
           & 21  &        & $<198$ \\\hline
\multirow{2}{*}{2022-08-28} & 5.5 & \multirow[c]{2}{*}{6D}     & $<53.7$\\
           & 9   &        & $<33.6$\\\hline
\multirow{2}{*}{2022-09-21} & 5.5 & \multirow[c]{2}{*}{6D}     & $<82.2$\\
           & 9   &        & $<67.2$\\\hline
\multirow{2}{*}{2022-09-29} & 5.5 & \multirow[c]{2}{*}{6D}     & $<41.1$\\
           & 9   &        & $<34.8$\\\hline
\multirow{2}{*}{2022-10-02} & 5.5 & \multirow[c]{2}{*}{6D}     & $<57.3$\\
           & 9   &        & $<43.5$\\\hline
\multirow{2}{*}{2022-10-05} & 5.5 & \multirow[c]{2}{*}{6D}     & $<46.5$\\
           & 9   &        & $75\pm10$\\\hline
\multirow{2}{*}{2022-10-07} & 5.5 & \multirow[c]{2}{*}{6D}     & $47\pm5$\\
           & 9   &        & $<40.3$\\\hline
\multirow{2}{*}{2022-10-15} & 5.5 & \multirow[c]{2}{*}{H214}     & $<243$\\
           & 9   &        & $<140$\\\hline
\multirow{2}{*}{2023-03-14} & 5.5 & \multirow[c]{2}{*}{750C}     & $<114$\\
           & 9   &        & $<66$\\\hline
\multirow{2}{*}{2023-03-31} & 5.5 & \multirow[c]{2}{*}{750C}     & $<76.5$\\
           & 9   &        & $<57$\\\hline
\hline
2022-10-02, & \multirow{3}{*}{5.5} & \multirow{3}{*}{6D} & \multirow{3}{*}{$35\pm5$}\\
2022-10-05, & & & \\
2022-10-07  & & & \\

2022-10-02,  &  \multirow{3}{*}{9} &  \multirow{3}{*}{6D} &  \multirow{3}{*}{$28\pm5$}\\
2022-10-05,  & & & \\
2022-10-07   & & & \\\hline

2023-03-14,& \multirow{2}{*}{5.5} & \multirow{2}{*}{750C} & \multirow{2}{*}{$51\pm13$} \\
2023-03-31 & & & \\
2023-03-14,& \multirow{2}{*}{9} & \multirow{2}{*}{750C} & \multirow{2}{*}{$<58$} \\
2023-03-31 & & & \\\hline

\hline
\end{tabular}
\tablefoot{Upper limits are reported at 3$\sigma$.}
\end{table}

The details of the X-ray observations and the X-ray spectral fitting results are listed in Table~\ref{tab:logxray}.

\clearpage
\onecolumn
\setlength\tabcolsep{2.7ex}
\begin{longtable}{cccccccc}
\caption{Log of X-ray observations.}\label{tab:logxray}\\
\hline\hline
MJD & $T_\text{exp}$ & $T_\text{in}/\Gamma$ & \sflux & ObsID & Mission & X-ray phase & cycle\\
\hline
\endfirsthead
\caption{continued.}\\
\hline\hline
MJD & $T_\text{exp}$ & $T_\text{in}/\Gamma$ & \sflux & ObsID & Mission & X-ray phase & Cycle\\
\hline
\endhead
\hline
\endfoot
58919.15 &    325 &                      3.0 &                    $<0.22$ &      eRASS1 &               eROSITA &  ---      & ---    \\[0.5mm]
59101.73 &    340 &         $64_{-18}^{+29}$ &     $0.51_{-0.26}^{+0.51}$ &      eRASS2$^\text{a}$ &               eROSITA &  \fxrise  & 1    \\[0.5mm]
59272.63 &    194 &      $2.5_{-0.2}^{+0.2}$ &     $7.76_{-0.76}^{+0.84}$ &      eRASS3 &                   eROSITA &  \fxplat  & 1    \\[0.5mm]
59281.86 &    123 &      $2.8_{-0.2}^{+0.2}$ &    $11.30_{-1.40}^{+1.61}$ &      eRASS3 &                   eROSITA &  \fxplat  & 1    \\[0.5mm]
59284.89 &   1683 &      $2.5_{-0.1}^{+0.1}$ &    $11.64_{-1.45}^{+1.66}$ & 00014135001 &           \mission{Swift} &  \fxplat  & 1    \\[0.5mm]
59290.94 &   1664 &                      3.0 &                    $<0.19$ &  4604010101 &           \mission{NICER} &  \fxdrop  & 1    \\[0.5mm]
59291.53 &   9941 &                      3.0 &                    $<0.13$ &  4604010102 &           \mission{NICER} &  \fxdrop  & 1    \\[0.5mm]
59292.21 &   6394 &                      3.0 &                    $<0.14$ &  4604010103 &           \mission{NICER} &  \fxdrop  & 1    \\[0.5mm]
59298.15 &   4387 &                      3.0 &                    $<0.18$ &  4604010201 &           \mission{NICER} &  \fxdrop  & 1    \\[0.5mm]
59300.20 &  10040 &        $99_{-40}^{+120}$ &     $0.03_{-0.02}^{+0.06}$ &  0862770201 &      \mission{XMM-Newton} &  \fxdrop  & 1    \\[0.5mm]
59305.09 &   2735 &                      3.0 &                    $<0.20$ &  4604010301 &           \mission{NICER} &  \fxfaint & 1    \\[0.5mm]
59320.20 &   3254 &                      3.0 &                    $<0.28$ & 00014135002 &           \mission{Swift} &  \fxfaint & 1    \\[0.5mm]
59322.59 &   1808 &                      3.0 &                    $<0.30$ &  4604010501 &           \mission{NICER} &  \fxfaint & 1    \\[0.5mm]
59323.49 &   4525 &                      3.0 &                    $<0.21$ &  4604010502 &           \mission{NICER} &  \fxfaint & 1    \\[0.5mm]
59325.27 &    230 &                      3.0 &                    $<0.35$ &  4604010504 &           \mission{NICER} &  \fxfaint & 1    \\[0.5mm]
59328.31 &    420 &                      3.0 &                    $<0.24$ &  4604010507 &           \mission{NICER} &  \fxfaint & 1    \\[0.5mm]
59329.73 &   3342 &                      3.0 &                    $<0.24$ & 00014135003 &           \mission{Swift} &  \fxfaint & 1    \\[0.5mm]
59340.45 &   1402 &                      3.0 &                    $<0.54$ & 00014135004 &           \mission{Swift} &  \fxfaint & 1    \\[0.5mm]
59391.16 &   7669 &                       78 &     $0.18_{-0.11}^{+0.18}$ & 00014135005$^\text{a}$ &       \mission{Swift} &  \fxrise  & 2    \\[0.5mm]
59399.00 &   6406 &      $4.3_{-1.1}^{+1.3}$ &     $0.47_{-0.27}^{+0.70}$ & 00014135006$^\text{a}$ &       \mission{Swift} &  \fxrise  & 2    \\[0.5mm]
59405.13 &   1793 &      $5.1_{-1.6}^{+1.9}$ &     $1.65_{-1.14}^{+4.34}$ & 00014135007$^\text{a}$ &       \mission{Swift} &  \fxrise  & 2    \\[0.5mm]
59413.03 &   1641 &      $2.6_{-0.7}^{+0.7}$ &     $0.68_{-0.29}^{+0.53}$ & 00014135008$^\text{a}$ &       \mission{Swift} &  \fxrise  & 2    \\[0.5mm]
59422.51 &   4304 &      $3.5_{-0.1}^{+0.1}$ &     $2.45_{-0.16}^{+0.21}$ &  4595020102$^\text{a,b}$ &     \mission{NICER} &  \fxrise  & 2    \\[0.5mm]
59423.35 &   2589 &      $3.5_{-0.1}^{+0.1}$ &     $2.45_{-0.16}^{+0.21}$ &  4595020103$^\text{a,b}$ &     \mission{NICER} &  \fxrise  & 2    \\[0.5mm]
59427.10 &   1631 &      $2.9_{-0.5}^{+0.6}$ &     $1.21_{-0.45}^{+0.74}$ & 00014135009$^\text{a}$ &       \mission{Swift} &  \fxrise  & 2    \\[0.5mm]
59428.95 &    788 &      $3.6_{-0.5}^{+0.2}$ &     $3.05_{-0.91}^{+0.84}$ &  4595020105$^\text{a}$ &       \mission{NICER} &  \fxrise  & 2    \\[0.5mm]
59432.96 &    764 &      $3.7_{-0.2}^{+0.3}$ &     $3.37_{-0.64}^{+1.00}$ &  4595020107$^\text{a,b}$ &     \mission{NICER} &  \fxrise  & 2    \\[0.5mm]
59433.12 &    999 &      $3.7_{-0.2}^{+0.3}$ &     $3.37_{-0.64}^{+1.00}$ &  4595020108$^\text{a,b}$ &     \mission{NICER} &  \fxrise  & 2    \\[0.5mm]
59434.71 &   1968 &      $3.5_{-0.5}^{+0.5}$ &     $3.07_{-0.96}^{+1.45}$ & 00014135010$^\text{a}$ &       \mission{Swift} &  \fxrise  & 2    \\[0.5mm]
59441.95 &    514 &      $2.9_{-0.1}^{+0.1}$ &     $4.55_{-0.32}^{+0.28}$ &  4595020109$^\text{a,b}$ &     \mission{NICER} &  \fxrise  & 2    \\[0.5mm]
59442.05 &    224 &      $2.9_{-0.1}^{+0.1}$ &     $4.55_{-0.32}^{+0.28}$ &  4595020110$^\text{a,b}$ &     \mission{NICER} &  \fxrise  & 2    \\[0.5mm]
59445.50 &   2121 &      $3.2_{-0.1}^{+0.1}$ &     $6.15_{-0.18}^{+0.19}$ &  4595020111$^\text{a}$ &       \mission{NICER} &  \fxrise  & 2    \\[0.5mm]
59447.51 &  40200 &                      --- &     $4.79_{-0.90}^{+1.10}$ &  0891801101 &      \mission{XMM-Newton} &  \fxplat  & 2    \\[0.5mm]
59448.38 &   1179 &      $3.2_{-0.1}^{+0.1}$ &     $5.06_{-0.21}^{+0.23}$ &  4595020112 &           \mission{NICER} &  \fxplat  & 2    \\[0.5mm]
59450.80 &   3350 &      $2.9_{-0.0}^{+0.0}$ &     $4.48_{-0.13}^{+0.13}$ &  4595020113 &           \mission{NICER} &  \fxplat  & 2    \\[0.5mm]
59453.89 &   2954 &      $2.8_{-0.2}^{+0.2}$ &     $3.20_{-0.59}^{+0.72}$ & 00014135011 &           \mission{Swift} &  \fxplat  & 2    \\[0.5mm]
59460.31 &    345 &      $3.4_{-0.2}^{+0.2}$ &     $5.26_{-0.67}^{+0.76}$ &      eRASS4 &                   eROSITA &  \fxplat  & 2    \\[0.5mm]
59467.45 &   1568 &      $2.8_{-0.3}^{+0.3}$ &     $2.39_{-0.29}^{+0.42}$ &  4604010901 &           \mission{NICER} &  \fxplat  & 2    \\[0.5mm]
59468.45 &    445 &      $2.9_{-0.9}^{+0.9}$ &     $1.73_{-0.91}^{+2.05}$ & 00014135012 &           \mission{Swift} &  \fxplat  & 2    \\[0.5mm]
59470.50 &    341 &      $4.1_{-0.9}^{+1.1}$ &     $1.67_{-0.61}^{+1.21}$ &  4604010902 &           \mission{NICER} &  \fxplat  & 2    \\[0.5mm]
59472.51 &   2542 &      $2.6_{-0.4}^{+0.4}$ &     $1.71_{-0.44}^{+0.61}$ & 00014135013 &           \mission{Swift} &  \fxplat  & 2    \\[0.5mm]
59476.22 &   1560 &      $2.9_{-0.4}^{+0.4}$ &     $1.23_{-0.17}^{+0.24}$ &  4604010903 &           \mission{NICER} &  \fxplat  & 2    \\[0.5mm]
59479.33 &    929 &      $3.0_{-0.3}^{+0.3}$ &     $2.81_{-0.35}^{+0.49}$ &  4604010904 &           \mission{NICER} &  \fxplat  & 2    \\[0.5mm]
59481.30 &  85960 &                      --- &     $2.40_{-0.36}^{+0.76}$ &  0891801701 &      \mission{XMM-Newton} &  \fxplat  & 2    \\[0.5mm]
59481.49 &   2490 &      $2.7_{-0.4}^{+0.4}$ &     $1.48_{-0.41}^{+0.58}$ & 00014135014 &           \mission{Swift} &  \fxplat  & 2    \\[0.5mm]
59482.46 &   4637 &      $2.9_{-0.3}^{+0.3}$ &     $2.33_{-0.26}^{+0.37}$ &  4604010905 &           \mission{NICER} &  \fxplat  & 2    \\[0.5mm]
59485.04 &   1183 &      $2.7_{-0.3}^{+0.3}$ &     $2.03_{-0.23}^{+0.31}$ &  4604010906 &           \mission{NICER} &  \fxplat  & 2    \\[0.5mm]
59488.08 &   1532 &      $3.0_{-0.3}^{+0.3}$ &     $2.78_{-0.38}^{+0.57}$ &  4604010907 &           \mission{NICER} &  \fxplat  & 2    \\[0.5mm]
59491.05 &   1410 &      $2.7_{-0.3}^{+0.3}$ &     $2.94_{-0.34}^{+0.49}$ &  4604010908 &           \mission{NICER} &  \fxplat  & 2    \\[0.5mm]
59494.25 &    811 &      $2.5_{-0.3}^{+0.3}$ &     $2.34_{-0.24}^{+0.31}$ &  4604010909 &           \mission{NICER} &  \fxplat  & 2    \\[0.5mm]
59496.67 &   3079 &      $2.7_{-0.3}^{+0.3}$ &     $2.05_{-0.49}^{+0.64}$ & 00014135016 &           \mission{Swift} &  \fxplat  & 2    \\[0.5mm]
59497.25 &   1133 &      $2.9_{-0.3}^{+0.3}$ &     $2.63_{-0.31}^{+0.44}$ &  4604010910 &           \mission{NICER} &  \fxplat  & 2    \\[0.5mm]
59499.80 &   1703 &      $2.5_{-0.3}^{+0.3}$ &     $2.19_{-0.24}^{+0.31}$ &  4604010911 &           \mission{NICER} &  \fxplat  & 2    \\[0.5mm]
59500.64 &   3647 &      $2.9_{-0.3}^{+0.3}$ &     $2.47_{-0.28}^{+0.41}$ &  4604010912 &           \mission{NICER} &  \fxplat  & 2    \\[0.5mm]
59501.35 &  10186 &      $2.9_{-0.2}^{+0.2}$ &     $2.13_{-0.22}^{+0.31}$ &  4604010913 &           \mission{NICER} &  \fxplat  & 2    \\[0.5mm]
59502.49 &   7885 &      $2.7_{-0.3}^{+0.3}$ &     $1.96_{-0.18}^{+0.24}$ &  4604010914 &           \mission{NICER} &  \fxplat  & 2    \\[0.5mm]
59503.48 &   7906 &      $2.9_{-0.2}^{+0.2}$ &     $2.48_{-0.24}^{+0.34}$ &  4604010915 &           \mission{NICER} &  \fxplat  & 2    \\[0.5mm]
59504.51 &   4370 &      $2.8_{-0.3}^{+0.3}$ &     $2.23_{-0.24}^{+0.35}$ &  4604010916 &           \mission{NICER} &  \fxplat  & 2    \\[0.5mm]
59505.52 &   3600 &      $2.7_{-0.3}^{+0.3}$ &     $2.36_{-0.25}^{+0.35}$ &  4604010917 &           \mission{NICER} &  \fxplat  & 2    \\[0.5mm]
59506.22 &    826 &      $2.8_{-0.3}^{+0.3}$ &     $2.41_{-0.29}^{+0.40}$ &  4604010918 &           \mission{NICER} &  \fxplat  & 2    \\[0.5mm]
59508.15 &    553 &      $3.3_{-0.5}^{+0.5}$ &     $1.95_{-0.39}^{+0.60}$ &  4604010920 &           \mission{NICER} &  \fxplat  & 2    \\[0.5mm]
59510.21 &   3079 &      $2.8_{-0.3}^{+0.3}$ &     $2.31_{-0.26}^{+0.38}$ &  4604010921 &           \mission{NICER} &  \fxplat  & 2    \\[0.5mm]
59510.83 &   2777 &      $2.3_{-0.3}^{+0.3}$ &     $1.71_{-0.39}^{+0.51}$ & 00014135017 &           \mission{Swift} &  \fxplat  & 2    \\[0.5mm]
59511.91 &   2492 &      $2.9_{-0.3}^{+0.3}$ &     $1.91_{-0.20}^{+0.27}$ &  4604010922 &           \mission{NICER} &  \fxplat  & 2    \\[0.5mm]
59517.60 &   2745 &                      3.0 &                    $<0.15$ &  4604011001 &           \mission{NICER} &  \fxdrop  & 2    \\[0.5mm]
59520.47 &   3000 &                      3.0 &                    $<0.22$ &  4604011002 &           \mission{NICER} &  \fxfaint & 2    \\[0.5mm]
59523.54 &   1200 &                      3.0 &                    $<0.27$ &  4604011003 &           \mission{NICER} &  \fxfaint & 2    \\[0.5mm]
59524.53 &   1371 &                      3.0 &                    $<1.50$ & 00014135018 &           \mission{Swift} &  \fxfaint & 2    \\[0.5mm]
59526.63 &    936 &                      3.0 &                    $<0.29$ &  4604011004 &           \mission{NICER} &  \fxfaint & 2    \\[0.5mm]
59527.63 &   3924 &                      3.0 &                    $<0.24$ & 00014135019 &           \mission{Swift} &  \fxfaint & 2    \\[0.5mm]
59529.83 &    636 &                      3.0 &                    $<0.30$ &  4604011005 &           \mission{NICER} &  \fxfaint & 2    \\[0.5mm]
59529.87 &   1402 &                      3.0 &                    $<0.69$ & 00014135021 &           \mission{Swift} &  \fxfaint & 2    \\[0.5mm]
59532.54 &   1143 &                      3.0 &                    $<0.29$ &  4604011006 &           \mission{NICER} &  \fxfaint & 2    \\[0.5mm]
59535.32 &   1817 &                      3.0 &                    $<0.28$ &  4604011007 &           \mission{NICER} &  \fxfaint & 2    \\[0.5mm]
59538.23 &   2041 &                      3.0 &                    $<0.19$ &  4604011008 &           \mission{NICER} &  \fxfaint & 2    \\[0.5mm]
59541.33 &   1884 &                      3.0 &                    $<0.14$ &  4604011009 &           \mission{NICER} &  \fxfaint & 2    \\[0.5mm]
59543.75 &   1996 &                      3.0 &                    $<0.39$ & 00014135022 &           \mission{Swift} &  \fxfaint & 2    \\[0.5mm]
59557.41 &   2372 &                      3.0 &                    $<0.41$ & 00014135023 &           \mission{Swift} &  \fxfaint & 2    \\[0.5mm]
59562.50 &    193 &                      3.0 &                    $<3.60$ & 00014135024 &           \mission{Swift} &  \fxfaint & 2    \\[0.5mm]
59566.74 &    614 &                      3.0 &                    $<1.60$ & 00014135025 &           \mission{Swift} &  \fxfaint & 2    \\[0.5mm]
59572.58 &   2635 &                      3.0 &                    $<0.38$ & 00014135026 &           \mission{Swift} &  \fxfaint & 2    \\[0.5mm]
59584.90 &   2886 &                      3.0 &                    $<0.39$ & 00014135027 &           \mission{Swift} &  \fxfaint & 2    \\[0.5mm]
59601.40 &    819 &                      180 &                    $<0.27$ &  4595020114 &           \mission{NICER} &  \fxrise  & 3    \\[0.5mm]
59602.56 &   3132 &                      180 &                    $<0.05$ &  4595020115$^\text{a}$ &       \mission{NICER} &  \fxrise  & 3    \\[0.5mm]
59603.21 &   1684 &                      180 &                    $<0.11$ &  4595020116 &           \mission{NICER} &  \fxrise  & 3    \\[0.5mm]
59604.76 &    432 &                      120 &                    $<0.21$ &  4595020117 &           \mission{NICER} &  \fxrise  & 3    \\[0.5mm]
59605.47 &    425 &                      120 &                    $<0.37$ &  4595020118 &           \mission{NICER} &  \fxrise  & 3    \\[0.5mm]
59607.27 &    765 &                      120 &                    $<0.19$ &  4595020119 &           \mission{NICER} &  \fxrise  & 3    \\[0.5mm]
59608.60 &   2918 &                      120 &                    $<0.08$ &  4595020120 &           \mission{NICER} &  \fxrise  & 3    \\[0.5mm]
59609.72 &   1146 &                       80 &                    $<0.23$ &  4595020121 &           \mission{NICER} &  \fxrise  & 3    \\[0.5mm]
59610.47 &   2152 &                       80 &                    $<0.13$ &  4595020122 &           \mission{NICER} &  \fxrise  & 3    \\[0.5mm]
59611.79 &    754 &                       80 &                    $<0.29$ &  4595020123 &           \mission{NICER} &  \fxrise  & 3    \\[0.5mm]
59612.47 &    758 &                       80 &                    $<0.08$ &  4595020124$^\text{a}$ &       \mission{NICER} &  \fxrise  & 3    \\[0.5mm]
59613.37 &   1443 &                       80 &                    $<0.19$ &  4595020125 &           \mission{NICER} &  \fxrise  & 3    \\[0.5mm]
59619.47 &   5808 &         $78_{-18}^{+24}$ &     $0.23_{-0.05}^{+0.12}$ &  4595020126$^\text{a}$ &       \mission{NICER} &  \fxrise  & 3    \\[0.5mm]
59625.99 &    787 &      $3.5_{-0.2}^{+0.3}$ &     $0.67_{-0.09}^{+0.11}$ &  4595020127$^\text{a,b}$ &     \mission{NICER} &  \fxrise  & 3    \\[0.5mm]
59626.06 &   1584 &      $3.5_{-0.2}^{+0.3}$ &     $0.67_{-0.09}^{+0.11}$ &  4595020128$^\text{a,b}$ &     \mission{NICER} &  \fxrise  & 3    \\[0.5mm]
59633.19 &   3669 &      $3.6_{-0.2}^{+0.2}$ &     $1.11_{-0.11}^{+0.11}$ &  4595020129$^\text{a}$ &       \mission{NICER} &  \fxrise  & 3    \\[0.5mm]
59635.13 &  10129 &      $3.2_{-0.3}^{+0.3}$ &     $0.77_{-0.20}^{+0.27}$ & 00014135099$^\text{a}$ &       \mission{Swift} &  \fxrise  & 3    \\[0.5mm]
59640.03 &   3707 &      $3.1_{-0.1}^{+0.1}$ &     $1.43_{-0.10}^{+0.10}$ &  4604010104$^\text{a}$ &       \mission{NICER} &  \fxrise  & 3    \\[0.5mm]
59640.03 &   3707 &      $3.1_{-0.1}^{+0.1}$ &     $1.43_{-0.10}^{+0.10}$ &  4604010105$^\text{a}$ &       \mission{NICER} &  \fxrise  & 3    \\[0.5mm]
59642.13 &   4069 &      $3.2_{-0.1}^{+0.1}$ &     $1.40_{-0.13}^{+0.08}$ &  4604010106 &           \mission{NICER} &  \fxplat  & 3    \\[0.5mm]
59643.68 &   2485 &      $3.2_{-0.5}^{+0.5}$ &     $1.78_{-0.58}^{+0.88}$ & 00014135033 &           \mission{Swift} &  \fxplat  & 3    \\[0.5mm]
59645.48 &   2460 &      $3.2_{-0.3}^{+0.3}$ &     $1.71_{-0.26}^{+0.38}$ &  4604010107 &           \mission{NICER} &  \fxplat  & 3    \\[0.5mm]
59648.20 &   1991 &      $3.2_{-0.4}^{+0.4}$ &     $1.54_{-0.24}^{+0.36}$ &  4604010108 &           \mission{NICER} &  \fxplat  & 3    \\[0.5mm]
59650.19 &   2977 &      $3.2_{-0.5}^{+0.5}$ &     $1.58_{-0.56}^{+0.88}$ & 00014135034 &           \mission{Swift} &  \fxplat  & 3    \\[0.5mm]
59651.90 &    668 &      $2.6_{-0.6}^{+0.6}$ &     $1.39_{-0.24}^{+0.35}$ &  4604010109 &           \mission{NICER} &  \fxplat  & 3    \\[0.5mm]
59652.06 &    720 &      $3.6_{-0.5}^{+0.6}$ &     $2.12_{-0.52}^{+0.91}$ &  4604010110 &           \mission{NICER} &  \fxplat  & 3    \\[0.5mm]
59654.83 &   1787 &      $3.7_{-0.4}^{+0.4}$ &     $1.83_{-0.37}^{+0.59}$ &  4604010111 &           \mission{NICER} &  \fxplat  & 3    \\[0.5mm]
59656.16 &   2665 &      $3.0_{-0.4}^{+0.4}$ &     $1.48_{-0.47}^{+0.68}$ & 00014135036 &           \mission{Swift} &  \fxplat  & 3    \\[0.5mm]
59657.15 &    --- &                      --- &     $1.92_{-0.04}^{+0.04}$ &  0884960601 &      \mission{XMM-Newton} &  \fxplat  & 3    \\[0.5mm]
59663.82 &   3172 &      $2.8_{-0.4}^{+0.4}$ &     $1.52_{-0.42}^{+0.58}$ & 00014135037 &           \mission{Swift} &  \fxplat  & 3    \\[0.5mm]
59670.56 &   3127 &      $2.5_{-0.4}^{+0.4}$ &     $1.22_{-0.32}^{+0.44}$ & 00014135038 &           \mission{Swift} &  \fxplat  & 3    \\[0.5mm]
59678.08 &   4305 &      $2.8_{-0.4}^{+0.5}$ &     $0.68_{-0.22}^{+0.34}$ & 00014135039 &           \mission{Swift} &  \fxplat  & 3    \\[0.5mm]
59684.37 &   2922 &      $2.6_{-0.5}^{+0.5}$ &     $0.88_{-0.29}^{+0.44}$ & 00014135040 &           \mission{Swift} &  \fxplat  & 3    \\[0.5mm]
59691.98 &   6391 &      $3.1_{-0.3}^{+0.4}$ &     $1.08_{-0.28}^{+0.39}$ & 00014135041 &           \mission{Swift} &  \fxplat  & 3    \\[0.5mm]
59698.54 &   2627 &      $3.5_{-0.8}^{+0.8}$ &     $1.18_{-0.56}^{+1.06}$ & 00014135042 &           \mission{Swift} &  \fxplat  & 3    \\[0.5mm]
59705.30 &   3014 &     $3.0_{-3.0}^{+-3.0}$ &     $0.12_{-0.08}^{+0.13}$ & 00014135043 &           \mission{Swift} &  \fxdrop  & 3    \\[0.5mm]
59752.82 &   1459 &                      3.0 &                    $<0.50$ & 00014135044 &           \mission{Swift} &  \fxfaint & 3    \\[0.5mm]
59756.64 &   2618 &                      3.0 &                    $<0.28$ & 00014135045 &           \mission{Swift} &  \fxfaint & 3    \\[0.5mm]
59759.19 &   3092 &                      3.0 &                    $<0.25$ & 00014135046 &           \mission{Swift} &  \fxfaint & 3    \\[0.5mm]
59766.09 &   2668 &                      3.0 &                    $<0.31$ & 00014135047 &           \mission{Swift} &  \fxfaint & 3    \\[0.5mm]
59773.49 &   2680 &                      3.0 &                    $<0.43$ & 00014135048 &           \mission{Swift} &  \fxfaint & 3    \\[0.5mm]
59780.34 &   3377 &                      3.0 &                    $<0.25$ & 00014135049 &           \mission{Swift} &  \fxrise  & 4    \\[0.5mm]
59787.41 &   2976 &                      3.0 &                    $<0.40$ & 00014135050 &           \mission{Swift} &  \fxrise  & 4    \\[0.5mm]
59794.63 &   2720 &                      3.0 &                    $<0.51$ & 00014135051 &           \mission{Swift} &  \fxrise  & 4    \\[0.5mm]
59801.23 &   4023 &      $3.1_{-2.2}^{+2.2}$ &     $0.12_{-0.09}^{+0.46}$ & 00014135052$^\text{a}$ &       \mission{Swift} &  \fxrise  & 4    \\[0.5mm]
59806.58 &   2627 & $3.0_{-102.0}^{+-102.0}$ &                    $<0.15$ &  5202950108$^\text{a}$ &       \mission{NICER} &  \fxrise  & 4    \\[0.5mm]
59811.17 &    187 &      $3.5_{-0.9}^{+1.3}$ &     $0.77_{-0.39}^{+0.41}$ &  5202950109$^\text{a}$ &       \mission{NICER} &  \fxrise  & 4    \\[0.5mm]
59811.48 &   3179 &      $2.3_{-0.9}^{+0.9}$ &     $0.24_{-0.12}^{+0.24}$ & 00014135053$^\text{a}$ &       \mission{Swift} &  \fxrise  & 4    \\[0.5mm]
59813.97 &    366 &      $3.4_{-0.7}^{+0.8}$ &     $0.55_{-0.24}^{+0.27}$ &  5202950110$^\text{a}$ &       \mission{NICER} &  \fxrise  & 4    \\[0.5mm]
59814.03 &    256 &      $2.8_{-0.6}^{+0.7}$ &     $0.53_{-0.18}^{+0.23}$ &  5202950111$^\text{a}$ &       \mission{NICER} &  \fxrise  & 4    \\[0.5mm]
59814.12 &   9662 &   $3.17_{-0.06}^{+0.06}$ &     $0.70_{-0.03}^{+0.03}$ &  0884962601$^\text{a}$ &  \mission{XMM-Newton} &  \fxrise  & 4    \\[0.5mm]
59819.86 &   2570 &      $2.4_{-0.6}^{+0.7}$ &     $0.43_{-0.19}^{+0.33}$ & 00014135455 &           \mission{Swift} &  \fxplat  & 4    \\[0.5mm]
59825.17 &   3384 &      $2.7_{-0.7}^{+0.7}$ &     $0.38_{-0.18}^{+0.33}$ & 00014135056 &           \mission{Swift} &  \fxplat  & 4    \\[0.5mm]
59828.19 &   3303 &      $4.0_{-0.3}^{+0.4}$ &     $0.78_{-0.10}^{+0.17}$ &  5202950114 &           \mission{NICER} &  \fxplat  & 4    \\[0.5mm]
59832.46 &   3117 &      $2.7_{-0.5}^{+0.5}$ &     $0.68_{-0.24}^{+0.38}$ & 00014135057 &           \mission{Swift} &  \fxplat  & 4    \\[0.5mm]
59839.22 &   2647 &      $2.9_{-0.7}^{+0.7}$ &     $0.58_{-0.26}^{+0.49}$ & 00014135058 &           \mission{Swift} &  \fxplat  & 4    \\[0.5mm]
59839.53 &   2717 &      $3.0_{-0.2}^{+0.2}$ &     $0.63_{-0.07}^{+0.07}$ &  5202950115 &           \mission{NICER} &  \fxplat  & 4    \\[0.5mm]
59842.32 &   1380 &      $4.4_{-1.0}^{+1.8}$ &     $0.64_{-0.25}^{+0.42}$ &  5202950116 &           \mission{NICER} &  \fxplat  & 4    \\[0.5mm]
59845.39 &   1529 &      $3.6_{-0.5}^{+0.5}$ &     $0.62_{-0.18}^{+0.14}$ &  5202950117 &           \mission{NICER} &  \fxplat  & 4    \\[0.5mm]
59846.51 &   2605 &      $3.4_{-1.2}^{+1.2}$ &     $0.56_{-0.31}^{+0.82}$ & 00014135059 &           \mission{Swift} &  \fxplat  & 4    \\[0.5mm]
59847.94 &    841 &      $3.1_{-0.5}^{+0.6}$ &     $0.57_{-0.18}^{+0.19}$ &  5202950118 &           \mission{NICER} &  \fxplat  & 4    \\[0.5mm]
59848.10 &   2940 &      $2.8_{-0.3}^{+0.3}$ &     $0.65_{-0.20}^{+0.09}$ &  5202950119 &           \mission{NICER} &  \fxplat  & 4    \\[0.5mm]
59851.44 &   1555 &      $3.8_{-0.4}^{+0.4}$ &     $0.81_{-0.15}^{+0.20}$ &  5202950120 &           \mission{NICER} &  \fxplat  & 4    \\[0.5mm]
59852.96 &    341 &                    $3.0$ &                    $<0.73$ &  5202950121 &           \mission{NICER} &  \fxplat  & 4    \\[0.5mm]
59853.86 &   2640 &      $3.3_{-0.8}^{+0.9}$ &     $0.67_{-0.35}^{+0.77}$ & 00014135060 &           \mission{Swift} &  \fxplat  & 4    \\[0.5mm]
59857.19 &   2144 &      $3.7_{-0.3}^{+0.5}$ &     $0.75_{-0.13}^{+0.20}$ &  5202950123 &           \mission{NICER} &  \fxplat  & 4    \\[0.5mm]
59860.20 &   1766 &      $3.8_{-0.3}^{+0.5}$ &     $0.71_{-0.12}^{+0.19}$ &  5202950124 &           \mission{NICER} &  \fxplat  & 4    \\[0.5mm]
59860.82 &   2555 &      $3.4_{-0.7}^{+0.8}$ &     $0.82_{-0.39}^{+0.79}$ & 00014135061 &           \mission{Swift} &  \fxplat  & 4    \\[0.5mm]
59863.58 &   4048 &      $3.3_{-0.3}^{+0.2}$ &     $0.57_{-0.10}^{+0.06}$ &  5202950125 &           \mission{NICER} &  \fxplat  & 4    \\[0.5mm]
59867.30 &   2122 &      $3.5_{-0.5}^{+0.8}$ &     $0.57_{-0.17}^{+0.10}$ &  5202950127 &           \mission{NICER} &  \fxplat  & 4    \\[0.5mm]
59869.51 &    977 &      $4.1_{-0.5}^{+0.6}$ &     $0.83_{-0.20}^{+0.29}$ &  5202950128 &           \mission{NICER} &  \fxplat  & 4    \\[0.5mm]
59872.21 &    392 &                    $3.0$ &                    $<0.56$ &  5202950129 &           \mission{NICER} &  \fxplat  & 4    \\[0.5mm]
59874.88 &   2650 &      $2.8_{-0.7}^{+0.8}$ &     $0.42_{-0.20}^{+0.39}$ & 00014135062 &           \mission{Swift} &  \fxplat  & 4    \\[0.5mm]
59874.91 &   2416 &      $2.6_{-0.3}^{+0.3}$ &     $0.65_{-0.11}^{+0.17}$ &  5202950130 &           \mission{NICER} &  \fxplat  & 4    \\[0.5mm]
59875.49 &    316 &                      3.0 &                    $<0.33$ &  5202950131 &           \mission{NICER} &  \fxplat  & 4    \\[0.5mm]
59875.49 &   1211 &      $3.4_{-1.0}^{+1.4}$ &     $0.24_{-0.15}^{+0.15}$ &  5202950131 &           \mission{NICER} &  \fxplat  & 4    \\[0.5mm]
59876.52 &   1326 &      $3.9_{-0.6}^{+0.8}$ &     $0.44_{-0.12}^{+0.23}$ &  5202950132 &           \mission{NICER} &  \fxplat  & 4    \\[0.5mm]
59876.52 &    459 &      $3.7_{-0.9}^{+1.2}$ &     $0.49_{-0.18}^{+0.29}$ &  5202950132 &           \mission{NICER} &  \fxplat  & 4    \\[0.5mm]
59877.30 &   2150 &      $3.0_{-0.2}^{+0.3}$ &     $0.47_{-0.07}^{+0.08}$ &  5202950133 &           \mission{NICER} &  \fxplat  & 4    \\[0.5mm]
59877.30 &   1266 &      $3.1_{-0.4}^{+0.5}$ &     $0.53_{-0.10}^{+0.13}$ &  5202950133 &           \mission{NICER} &  \fxplat  & 4    \\[0.5mm]
59878.36 &   1955 &      $4.3_{-0.5}^{+0.6}$ &     $0.49_{-0.12}^{+0.17}$ &  5202950134 &           \mission{NICER} &  \fxdrop  & 4    \\[0.5mm]
59879.59 &   3164 &                    $3.0$ &                    $<0.55$ &  5202950135 &           \mission{NICER} &  \fxdrop  & 4    \\[0.5mm]
59879.59 &   1564 &      $4.1_{-0.6}^{+0.7}$ &     $0.43_{-0.12}^{+0.16}$ &  5202950135 &           \mission{NICER} &  \fxdrop  & 4    \\[0.5mm]
59881.55 &   1119 &      $2.3_{-0.7}^{+0.7}$ &     $0.33_{-0.07}^{+0.08}$ &  5202950136 &           \mission{NICER} &  \fxdrop  & 4    \\[0.5mm]
59881.55 &   3311 &      $1.4_{-1.4}^{+0.6}$ &     $0.17_{-0.09}^{+0.06}$ &  5202950136 &           \mission{NICER} &  \fxdrop  & 4    \\[0.5mm]
59882.80 &   8366 &     $3.0_{-3.0}^{+-3.0}$ &     $0.07_{-0.04}^{+0.06}$ &    14136364 &           \mission{Swift} &  \fxdrop  & 4    \\[0.5mm]
59884.59 &   3046 & $3.0_{-102.0}^{+-102.0}$ &                    $<0.11$ &  5202950137 &           \mission{NICER} &  \fxdrop  & 4    \\[0.5mm]
59884.59 &   2552 &                      3.0 &                    $<0.14$ &  5202950137 &           \mission{NICER} &  \fxdrop  & 4    \\[0.5mm]
59887.94 &   2868 &                      3.0 &                    $<0.36$ & 00014135065 &           \mission{Swift} &  \fxfaint & 4    \\[0.5mm]
59895.56 &   2688 &                      3.0 &                    $<0.35$ & 00014135066 &           \mission{Swift} &  \fxfaint & 4    \\[0.5mm]
59920.62 &   3455 &                      3.0 &                    $<0.31$ & 00014135067 &           \mission{Swift} &  \fxfaint & 4    \\[0.5mm]
59941.75 &   4054 &                      3.0 &                    $<0.32$ & 00014135068 &           \mission{Swift} &  \fxfaint & 4    \\[0.5mm]
59962.62 &   3984 &                      3.0 &                    $<0.23$ & 00014135069 &           \mission{Swift} &  \fxfaint & 4    \\[0.5mm]
59980.96 &   2963 &                      200 &                    $<0.25$ & 00014135070 &           \mission{Swift} &  \fxrise  & 5    \\[0.5mm]
59985.53 &  15381 &        $193_{-25}^{+43}$ &  $0.021_{-0.003}^{+0.005}$ &  0915390401$^\text{a}$ &  \mission{XMM-Newton} &  \fxrise  & 5    \\[0.5mm]
59990.59 &   2926 &                      100 &                    $<0.67$ & 00014135071 &           \mission{Swift} &  \fxrise  & 5    \\[0.5mm]
59999.79 &   3384 &      $3.4_{-2.3}^{+2.4}$ &     $0.26_{-0.17}^{+0.89}$ & 00014135072 &           \mission{Swift} &  \fxrise  & 5    \\[0.5mm]
60007.25 &  11300 &   $3.06_{-0.07}^{+0.07}$ &     $0.44_{-0.02}^{+0.03}$ &  0915390501$^\text{a}$ &  \mission{XMM-Newton} &  \fxrise  & 5    \\[0.5mm]
60009.45 &   2690 &      $3.2_{-1.3}^{+1.3}$ &     $0.53_{-0.27}^{+0.75}$ & 00014135073 &           \mission{Swift} &  \fxrise  & 5    \\[0.5mm]
60021.39 &   3189 &      $2.2_{-0.9}^{+0.9}$ &     $0.27_{-0.13}^{+0.25}$ & 00014135074 &           \mission{Swift} &  \fxplat  & 5    \\[0.5mm]
60023.00 &   9948 &      $2.6_{-0.4}^{+0.4}$ &     $0.29_{-0.12}^{+0.24}$ &       27737 &         \mission{Chandra} &  \fxplat  & 5    \\[0.5mm]
60031.16 &   2680 &      $3.2_{-0.8}^{+0.8}$ &     $0.63_{-0.30}^{+0.58}$ & 00014135075 &           \mission{Swift} &  \fxplat  & 5    \\[0.5mm]
60045.09 &  10059 &      $2.6_{-0.5}^{+0.5}$ &     $0.25_{-0.12}^{+0.27}$ &       27738 &         \mission{Chandra} &  \fxplat  & 5    \\[0.5mm]
60048.22 &   9948 &      $2.5_{-0.5}^{+0.5}$ &     $0.19_{-0.09}^{+0.20}$ &       27798 &         \mission{Chandra} &  \fxplat  & 5    \\[0.5mm]
60072.18 &   4795 &      $3.0_{-0.8}^{+0.9}$ &     $0.28_{-0.15}^{+0.32}$ & 00014135076 &           \mission{Swift} &  \fxplat  & 5    \\[0.5mm]
\hline
%\end{supertabular}
\end{longtable}
\tablefoot{\textit{MJD} is the mid-date of the coverage for each observation; \textit{$T_\text{exp}$} is the effective exposure time in units of seconds. Exposure for \mission{XMM-Newton} ObsID 0862770201 is calculated from combined MOS data, while the pn exposures are given for the other \mission{XMM-Newton} observations; \textit{$T_\text{in}/\Gamma$}: the inner temperature of a multi-color disk or the photon index of the power law. A value without an uncertainty means the parameter is fixed at the given value during spectral fitting; \textit{\sflux} is the rest frame intrinsic $0.2-2.0\,$keV flux in units of $10^{-12}\,\unitflux$. The $3\sigma$ upper limits are given for observations in which the \jsrc is not detected; \textit{ObsID} is the observation ID for each observation; \textit{Mission}: the telescopes of which the data were taken; \textit{X-ray phase} marks the phases of the \jsrc during the observations; \textit{Cycle}: the cycle the X-ray flare during which the data were taken. Quoted uncertainties are at the 90\% confidence level. Upper limits are reported at $3\sigma$.\\
a: observations used in modelling the profile of the \fxrise phase.\\
b: the X-ray spectral fitting results were from jointly fitting of the two consecutive observations.}
\clearpage
\twocolumn

\end{appendix}

\end{document}